\documentclass[twocolumn]{aastex631}
\shorttitle{Asymmetry of the activity line profiles of II~Peg}
\shortauthors{Cao \& Gu}
\graphicspath{{./}{figures/}}

\begin{document}

\title{Doppler-shifted Signatures in the Chromospheric Line Profiles of II~Pegasi}

\correspondingauthor{Dongtao Cao, Shenghong Gu}
\email{dtcao@ynao.ac.cn, shenghonggu@ynao.ac.cn}

\author[0000-0002-3534-1740]{Dongtao Cao}
\affiliation{Yunnan Observatories, Chinese Academy of Sciences, Kunming 650216, China}
\affiliation{Key Laboratory for the Structure and Evolution of Celestial Objects, Chinese Academy of Sciences, Kunming 650216, China}
\affiliation{International Centre of Supernovae, Yunnan Key Laboratory, Kunming 650216, China}

\author{Shenghong Gu}
\affiliation{Yunnan Observatories, Chinese Academy of Sciences, Kunming 650216, China}
\affiliation{Key Laboratory for the Structure and Evolution of Celestial Objects, Chinese Academy of Sciences, Kunming 650216, China}
\affiliation{School of Astronomy and Space Science, University of Chinese Academy of Sciences, Beijing 101408, China}


\begin{abstract}
Stellar flares are frequently accompanied by prominence eruptions, coronal mass ejections (CMEs), and other forms of plasma motion. Based on the long-term high-resolution spectroscopic monitoring data set of the RS CVn-type star II~Pegasi (II~Peg), we search for stellar prominence eruptions and potential CMEs by investigating asymmetric signatures in the H$_{\alpha}$ and H$_{\beta}$ line profiles during the flare events. We observed seven optical flares with asymmetric spectral profiles, releasing low limit energies of $10^{33}$--$10^{34}$~erg in the H$_{\alpha}$ line. Five of these flares were associated with six potential CMEs, identified through the Doppler-shifted emission signatures in the chromospheric line profiles, particularly the H$_{\alpha}$ line. The mass estimates for the CME candidates range from $10^{19}$ to $10^{20}$~g, while the kinetic energies are calculated to be around $10^{33}$--$10^{34}$~erg. The most massive CME candidate was simultaneously recorded in the H$_{\alpha}$, H$_{\beta}$, and $\mbox{He~{\sc i}}$~D$_{3}$ lines. For II~Peg, the possible flare-CME association rate is about 30\%. There were also some redshifted excess absorption signatures in the H$_{\alpha}$ line profiles. These signatures could not only be observed in the flaring spectra but also in the spectra without flare, which could be interpreted as features resulting from flare-driven coronal rain and quiescent rain, respectively. 
\end{abstract}

\keywords{Stellar activity (1580); Optical flares (1166); Stellar coronal mass ejections (1881); Stellar chromospheres (230); Stellar coronae (305); Spectroscopy (1558); RS Canum Venaticorum variable stars (1416)}
\section{Introduction} \label{sec1}
Stellar coronal mass ejections (CMEs), manifested by large expulsions of plasma and magnetic field into the interplanetary space, have attracted much attention in recent years. Frequently occurred CMEs may be an important contribution to stellar mass and angular momentum loss, and are therefore relevant to stellar evolution \citep{Aarnio2012, Osten2015}. On the other hand, with an increasing number of exoplanets discovered, it is necessary to characterize the effects of stellar magnetic activity on the planetary atmosphere and habitability. As the major driver of exoplanetary space weather, stellar CMEs from host stars may have severe impacts on the environments of surrounding exoplanets \citep{Airapetian2016, Cherenkov2017, Hazra2022, Cohen2022}.

While CMEs on the Sun have been frequently detected in the past few decades, with several CMEs per day during high activity phase \citep{Webb2012}, reported stellar CME detections have been rare. Because we are unable to directly image stellar CMEs with current instrumentation, as is done for the Sun, the detection is mainly based on some
other evidence. Possible stellar CMEs have been detected so far by using the Doppler-shifted emission or absorption signatures in the optical, UV and X-ray spectral lines \citep{Houdebine1990, Leitzinger2011,  Argiroffi2019, Namekata2022, Lu2022, Chen2022, Inoue2023}, X-ray, extreme-UV, and far-UV dimming \citep[e.g.][]{Veronig2021}, and X-ray continuous absorption \citep{Favata1999, Moschou2017}. A detailed review for the different methodologies of the detection of stellar CMEs can be found in \citet{Leitzinger2022} and \citet{Tian2023}. It is generally accepted that the possible stellar CMEs detected in the optical spectral line profiles mostly correspond to eruptive prominences that are believed to form the CME cores in the frequently observed three-part structure of solar CMEs \citep{Forbes2000}. If an erupting prominence lies directly on the line of sight to the observer, referred as a filament, it may produce an absorption signature \citep[see][]{Namekata2022, Cao2025}. 

In addition, the Doppler-shifted emission and absorption signatures observed in the chromospheric line profiles may also be attributed to other forms of plasma motion occurring during the flares; for example, in the form of chromospheric evaporation \citep{Tei2018}, or chromospheric condensation \citep{Ichimoto1984, Fuhrmeister2018, Vida2019}, or coronal rains \citep{Antolin2012, Lacatus2017, Fuhrmeister2018}. These possibility would be more prominent when the velocities of the Doppler-shifted signatures are smaller.

Since 1999, at Yunnan Observatories, our research group has begun a long-term high-resolution spectroscopic monitoring project for the very active RS~CVn-type star II~Peg to study its magnetic activity. Therefore, we have an opportunity to search for stellar CMEs or plasma motion in other forms by analyzing the line asymmetry features appeared in its chromospheric activity line profiles. Just as some research works have attempted to detect possible stellar CMEs by using long-term high-resolution spectroscopic observations of one star \citep{Muheki2020a, Muheki2020b} and investigating a large number of stars based on a large amount of archival data \citep[e.g.][]{Fuhrmeister2018, Vida2019, Leitzinger2020, Koller2021, Lu2022}. 

II~Peg is a single-lined spectroscopic binary system, consisting of a K2~IV primary star and an unseen companion with a period of about 6.7~days \citep{Berdyugina1998}. More information of physical parameters of II~Peg can be found in our previous paper (\citealt{Cao2024}, hereafter C24). As a member of RS CVn-type stars, the K2~IV primary star of II~Peg shows remarkable photometric starspot activity \citep[e.g.][]{Vogt1981, Berdyugina1998b, Rodono2000, Lindborg2013}, intense $\mbox{Ca~{\sc ii}}$~H\&K and H$_{\alpha}$ line emission \citep[e.g.][]{Vogt1981, Huenemoerder1987, Montes1997, Berdyugina1999, Frasca2008}, and strong UV and X-ray radiation \citep[e.g.][]{Byrne1989, Ercolano2008}. In the solar context, CMEs are frequently associated with flares, particularly in the cases of energetic flares \citep{Yashiro2006,Aarnio2011}. II~Peg is a high-rate flaring star and has shown many flare activities in the optical, UV, and X-ray wavelength bands \citep[e.g.][]{Rodono1987, Doyle1991, Berdyugina1999, Frasca2008, Ercolano2008, Siwak2010, Tsuboi2016}. Therefore, it can be hypothesized that II~Peg may have frequent CME occurrences based on the close association between solar CMEs and flares. More recently, a potential flare-associated CME candidate was identified by us on II~Peg in C24, showing a prominent redshifted emission signature in the high-resolution H$_{\alpha}$ spectra with a large bulk velocity of about 429~km~s$^{-1}$. The large velocity greatly exceeds the escape velocity of the K2~IV primary star of II~Peg.

In this study, based on long-term high-resolution spectroscopic monitoring data set of II~Peg, strong optical flare events are identified, among which we search for stellar prominence eruptions and potential CMEs by investigating asymmetric features in the H$_{\alpha}$ and H$_{\beta}$ line profiles. In Section~\ref{sec2}, we provide details of spectroscopic observations analyzed in this study, the data reduction, and the spectral subtraction of H$_{\alpha}$ and H$_{\beta}$ lines. Section~\ref{sec3} presents the detected optical flare events, whose spectral line profiles show asymmetric behavior. The Doppler-shifted emission signatures in the spectral line profiles during the flares are described in Section~\ref{sec4}. Section~\ref{sec5} presents the redshifted excess absorption signatures found in the line profiles. In Section~\ref{sec6}, a comprehensive discussion is provided. Finally, we conclude and summarize new results in Section~\ref{sec7}.
\section{Spectroscopic observations, data reduction, and spectral subtraction}\label{sec2}
The detailed observing information of high-resolution spectra analyzed here is listed in Table~\ref{tab1}, which includes observing date, exposure time, heliocentric Julian date (HJD), and orbital phase calculated with the ephemeris from \citet{Rosen2015}. The zero phase is defined as the time when the K2~IV primary star of II~Peg exhibits maximum positive radial velocity.

Before 2012, II~Peg was observed with the Coud\'{e} Echelle Spectrograph (CES, \citealt{Zhao2001}) mounted on the 2.16 m telescope at the Xinglong station, National Astronomical Observatories, Chinese Academy of Sciences, China. The spectrograph covers a wavelength range of about 5600--9000~\AA~with an average resolving power R~=~$\lambda$/$\Delta\lambda$~$\simeq$~37000, and the spectra were recorded on a $1024 \times 1024$ pixel Tektronix CCD detector. Therefore, the H$_{\beta}$ line is not covered in the CES spectra. Later, spectroscopic observations were performed with the fiber-fed high-resolution spectrograph (HRS, \citealt{Fan2016}) installed on the Xinglong 2.16 m telescope. HRS produces spectra with a resolving power R~=~$\lambda$/$\Delta\lambda$~$\simeq$~48000 in a wavelength range of 3900--10000~\AA, using a $4096~\times~4096$ pixel CCD detector. Besides II~Peg, observations of some rapidly rotating early-type stars and slowly rotating inactive stars were also obtained by using the same instrument configuration. The spectra of early-type stars were used as telluric templates, whereas the inactive stars were used as references in the following spectral subtraction.

The data reduction was performed with the IRAF\footnote{IRAF is distributed by the National Optical Astronomy Observatories, which is operated by the Association of Universities for Research in Astronomy (AURA), Inc., under cooperative agreement with the National Science Foundation.} package, following the procedure outlined in C24. In some of our observations, there were heavy telluric absorption lines in the activity line regions of interest, especially for the H$_{\alpha}$ and $\mbox{He~{\sc i}}$ D$_{3}$ lines. We eliminated them by using the spectra of two brighter and rapidly rotating early-type stars HR~8858 (B5~V, $vsini$ = 316~km~s$^{-1}$) and HR~7894 (B5~IV, $vsini$ = 285~km~s$^{-1}$) as telluric absorption line templates. This process was carried out through an interactive procedure in the IRAF package.
 
To separate the activity contribution from the observed H$_{\alpha}$ and H$_{\beta}$ line profiles, we apply the spectral subtraction technique utilizing the STARMOD program \citep{Barden1985, Montes1997, Montes2000}. This program can subtract a synthesized spectrum from the observed one. The synthesized spectrum is constructed through artificially rotationally broadening and radial velocity shifting for the spectrum of an inactive star (HD~3351, K0~IV), which shares the similar spectral type and luminosity class with the primary star of II~Peg. The rotational velocity ($vsini$) value of 21.6~km~s$^{-1}$ was used in the construction. Therefore, the subtraction between the observed and synthesized spectra represents the pure activity contribution of II~Peg. The activity contribution could be obtained by measuring the equivalent widths (EWs) of the subtracted H$_{\alpha}$ and H$_{\beta}$ line profiles. More detailed information on the spectral subtraction procedure can also be found in C24.
\begin{deluxetable}{cccc}
\tablenum{1}
\tablecaption{II~Peg's Observing Log\label{tab1}}
\tablewidth{0pt}
\tablehead{
\colhead{\bf{UT date}} &\colhead{\bf{Exp.time}} &\colhead{\bf{HJD}} &\colhead{\bf{Phase}}\\
\nocolhead{} & \colhead{\bf{(s)}} & \colhead{\bf{(2,450,000+)}} & \nocolhead{}
}
\startdata
\multicolumn{4}{c}{\bf{2.16 m telescope~+~HRS}} \\
2022 Dec 1 & 1800 & 9915.038 & 0.807\\
2022 Dec 1 & 1800 & 9915.121 & 0.819\\
2022 Dec 1 & 1800 & 9915.158 & 0.825\\
2022 Dec 3 & 2700 & 9916.996 & 0.098\\
2022 Dec 3 & 2700 & 9917.078 & 0.111\\
2022 Dec 4 & 1800 & 9917.970 & 0.243\\
2022 Dec 4 & 2700 & 9918.078 & 0.259\\
2022 Dec 4 & 2700 & 9918.125 & 0.266\\
\enddata
\tablecomments{Table~\ref{tab1} is published in its entirety in the machine-readable format. A portion is shown here for guidance regarding its form and content.}
\end{deluxetable}

\begin{deluxetable*}{ccccccccc}
\tablenum{2}
\tablecaption{Optical Flare Events\label{tab2}}
\tablewidth{10pt}
\tablehead{
\colhead{\bf{State}} &\colhead{\bf{UT date}} &\colhead{\bf{Phase}}&\colhead{\bf{EW$_{H\beta}$}}&\colhead{\bf{EW$_{H\alpha}$}}&\colhead{\bf{EW$_{H\beta}^{flare}$}}&\colhead{\bf{EW$_{H\alpha}^{flare}$}} &\colhead{\bf{E$_{H\alpha}^{flare}$}}&\colhead{\bf{E$_{WL, bol}^{flare}$}}\\
\nocolhead{}&\nocolhead{}&\nocolhead{}&\colhead{\bf{(\AA)}} &\colhead{\bf{(\AA)}}&\colhead{\bf{(\AA)}} &\colhead{\bf{(\AA)}}&\colhead{\bf{(erg)}}&\colhead{\bf{(erg)}}
}
\startdata
&&&&\bf{Flare \#1}\\
       Flaring  &2022 Dec 4&0.243   &1.692$\pm$0.030&3.845$\pm$0.046&1.153$\pm$0.023&1.778$\pm$0.046&$2.8\times10^{34}$&$1.5\times10^{36}$\\
       Flaring  &2022 Dec 4&0.259   &2.060$\pm$0.008&4.213$\pm$0.045&1.514$\pm$0.033&2.206$\pm$0.017\\
       Flaring  &2022 Dec 4&0.266   &1.868$\pm$0.015&4.060$\pm$0.047&1.318$\pm$0.026&2.064$\pm$0.016\\
       Preflare&\multicolumn{2}{c}{2022 Dec 1-3, 5, 6}&0.546$\pm$0.029&2.038$\pm$0.017&\nodata        &\nodata\\
\hline
&&&&\bf{Flare \#2}\\
       Flaring  &2019 Dec 11&0.304   &1.193$\pm$0.012&3.057$\pm$0.016&0.535$\pm$0.015&0.755$\pm$0.008&$6.1\times10^{33}$&$3.2\times10^{35}$\\
       Flaring  &2019 Dec 11&0.308   &1.115$\pm$0.010&2.914$\pm$0.016&0.455$\pm$0.022&0.620$\pm$0.057\\
       Preflare&\multicolumn{2}{c}{2019 Dec 3-9}&0.600$\pm$0.007&2.309$\pm$0.003&\nodata        &\nodata\\
\hline
&&&&\bf{Flare \#3}\\
       Flaring  &2019 Jan 16&0.361  &1.723$\pm$0.012&4.828$\pm$0.043&1.049$\pm$0.035&2.315$\pm$0.038&$1.4\times10^{34}$&$7.6\times10^{35}$\\
       Flaring  &2019 Jan 16&0.377  &1.444$\pm$0.022&4.453$\pm$0.053&0.720$\pm$0.027&1.919$\pm$0.020\\
       Preflare&\multicolumn{2}{c}{2019 Jan 15-17, 20, 21}&0.657$\pm$0.065&2.568$\pm$0.018&\nodata        &\nodata\\
\hline
&&&&\bf{Flare \#4}\\
       Flaring  &2017 Dec 5&0.831&1.318$\pm$0.006&3.706$\pm$0.018&0.602$\pm$0.033&1.341$\pm$0.015&$3.0\times10^{33}$&$1.5\times10^{35}$\\
       Preflare&2017 Dec 5&0.862&0.652$\pm$0.025&2.407$\pm$0.030&\nodata        &\nodata\\
\hline
&&&&\bf{Flare \#5}\\
       Flaring  &2006 Dec 9&0.898  &\nodata&3.290$\pm$0.023&\nodata&0.372$\pm$0.063&$1.5\times10^{34}$&$8.0\times10^{35}$\\
       Flaring  &2006 Dec 9&0.915  &\nodata&2.463$\pm$0.012&\nodata&-0.508$\pm$0.072\\
       Flaring  &2006 Dec 10&0.051  &\nodata&2.576$\pm$0.023&\nodata&-0.344$\pm$0.039\\
       Flaring  &2006 Dec 10&0.063  &\nodata&3.244$\pm$0.001&\nodata&0.300$\pm$0.049\\
       Flaring  &2006 Dec 11&0.195  &\nodata&3.607$\pm$0.016&\nodata&0.619$\pm$0.007\\
       Flaring  &2006 Dec 11&0.213  &\nodata&4.728$\pm$0.062&\nodata&1.779$\pm$0.048\\
       Preflare&\multicolumn{2}{c}{2006 Nov 29-30, Dec 6-8}&\nodata&2.952$\pm$0.010&\nodata&\nodata\\
\hline
&&&&\bf{Flare \#6}\\
       Flaring  &2005 Nov 22&0.107  &\nodata&3.611$\pm$0.069&\nodata&0.962$\pm$0.046&$2.6\times10^{34}$&$1.4\times10^{36}$\\
       Flaring  &2005 Nov 23&0.254  &\nodata&4.635$\pm$0.034&\nodata&1.929$\pm$0.010\\
       Flaring  &2005 Nov 24&0.407  &\nodata&3.720$\pm$0.029&\nodata&1.008$\pm$0.017\\
       Preflare&\multicolumn{2}{c}{2005 Nov 18, 20, 21}&\nodata&2.696$\pm$0.021&\nodata&\nodata\\
\hline
&&&&\bf{Flare \#7}\\
       Flaring  &1999 July 27&0.581  &\nodata&3.072$\pm$0.022&\nodata&0.266$\pm$0.041&$6.5\times10^{33}$&$3.4\times10^{35}$\\
       Flaring  &1999 July 27&0.584  &\nodata&3.200$\pm$0.039&\nodata&0.346$\pm$0.028\\
       Flaring  &1999 July 27&0.588  &\nodata&3.110$\pm$0.001&\nodata&0.334$\pm$0.051\\
       Flaring  &1999 July 27&0.591  &\nodata&3.140$\pm$0.020&\nodata&0.302$\pm$0.059\\
       Flaring  &1999 July 27&0.594  &\nodata&3.127$\pm$0.025&\nodata&0.326$\pm$0.069\\
       Flaring  &1999 July 27&0.597  &\nodata&3.165$\pm$0.002&\nodata&0.355$\pm$0.051\\
       Preflare&\multicolumn{2}{c}{1999 July 26-29, 31, Aug 1}&\nodata&2.842$\pm$0.032&\nodata&\nodata\\
\enddata
\tablecomments{EW$_{H\alpha}$ and EW$_{H\beta}$ represent the EWs of the STARMOD subtracted H$_{\alpha}$ and H$_{\beta}$ line profiles. EW$_{H\alpha}^{flare}$ and EW$_{H\beta}^{flare}$ quantify the differential spectra of the H$_{\alpha}$ and H$_{\beta}$ lines that are attributed to flare radiation. The term flaring state refers to the spectra observed during a flare, while the preflare state denotes either the mean spectrum or the spectrum recorded after the flare, as detailed in Sections 3 and 4.4. The EWs of the preflare spectra for both the H$_{\alpha}$ and H$_{\beta}$ lines are provided herein. Additionally, the dates of observations utilized to derive these preflare spectra (mean spectra) are also provided. The negative EWs indicate that the H$_{\alpha}$ line profile, after subtracting the preflare spectrum during observations conducted on 2006 December~9 and 10, exhibited a complete absorption feature. Or alternatively, the majority of the profile displayed an absorption characteristic. In the calculation of flare energy for each flare, all spectra identified as being in a flaring state are utilized.}
\end{deluxetable*}

\section{Optical flares and flare energy}\label{sec3}
In the optical wavelength range, the $\mbox{He~{\sc i}}$~D$_{3}$ line is a commonly used indicator to track flare activity in solar and stellar chromospheres, due to its very high excitation potential \citep[e.g.][]{Zirin1988, Gu2002,  Garcia2003, Cao2019}. During flare events, the $\mbox{He~{\sc i}}$~D$_{3}$ line typically exhibits a pronounced emission feature. Additionally, other chromospheric activity lines, such as H$_{\alpha}$ and H$_{\beta}$, also exhibit enhanced emission. Based on these observational characteristics, we are able to identify optical flare events within our long-term spectroscopic dataset of II~Peg. 

During the course of our observations, seven optical flares are identified, whose spectra exhibit asymmetric features, particularly for the H$_{\alpha}$ line. The observing information of these flares and the EWs of the subtracted lines are listed in Table~\ref{tab2}. The STARMOD subtracted spectra contain both intrinsic chromospheric emission and flare radiation. Typically, a preflare spectrum subtraction is performed to isolate pure flare radiation. However, in our case, since the closest preflare spectrum was obtained prior to the night, and the chromospheric line profiles of II~Peg may vary significantly from one night to another. Moreover, we did not obtain the preflare or post-flare spectra for each flare event taken at orbital phases close to that of the flare event. Therefore, we utilize a mean spectrum as the preflare reference spectrum to subtract from the observed flaring spectra, with the exception of observation taken on 2017 December~5. For each observing run, we build a mean spectrum by averaging all available spectra while excluding those captured during flare events and those exhibiting pronounced chromospheric activity. Additionally, the selected spectra do not display significant wing enhancements or absorptions. It is also important to note that the spectra collected over several days (see Table~\ref{tab2}), which correspond to approximately one orbital period, exhibited a relatively stable overall activity level during this short interval. Nevertheless, the rotational modulation of chromospheric active regions may still introduce a bias in each mean spectrum, potentially leading to inaccuracies in the estimation of EWs for flaring spectra and consequently affecting the assessment of flare energy. In contrast, for the observation on 2017 December~5, a distinct spectrum serves as the preflare reference (see Section 4.3). Subsequently, the EWs representing pure flare radiation are measured in the differential profiles of the H$_{\alpha}$ and H$_{\beta}$ lines. The resulting values are also cataloged in Table~\ref{tab2}.

We compute the stellar continuum flux $F_{H{\alpha}}$~(in erg cm$^{-2}$ s$^{-1}$ \AA$^{-1}$) near the H$_{\alpha}$ line region as a function of the color index $B-V$ ($\sim$~1.031 for II~Peg; \citealt{Messina2008}) based on the empirical relationship: 
\begin{eqnarray}
\log{F_{H{\alpha}}}=[7.538-1.081(B-V)]\pm{0.33}\nonumber \\
0.0~\leq~B-V~\leq~1.4
\end{eqnarray}
from \citet{Hall1996}, and then convert the EW representing the pure flare radiation into an absolute surface flux $F_{S}$~(in erg~cm$^{-2}$~s$^{-1}$). Therefore, the luminosity is calculated by using the derived absolute surface flux and the stellar radius $R_{\star}$~=~3.4$R_{\sun}$ of the K2~IV primary star of II~Peg \citep{Berdyugina1998}. 

For each flare, the luminosity of each flaring spectrum is integrated over the exposure time, and if multiple spectra are observed, they are summed together. This process yields an estimated rough flare energy (E$_{H\alpha}^{flare}$) of approximately \(10^{33}\)--\(10^{34}\) erg released in the H$_{\alpha}$ line, which serves as a lower limit for the flare energy. Furthermore, based on the relationship between H$_{\alpha}$ (E$_{H\alpha}^{flare}$) and bolometric white-light (E$_{WL, bol}^{flare}$) flare energies as described in Equation~(2) of \citet{Namekata2024}, we also derive that the bolometric white-light flare radiation energies E$_{WL, bol}^{flare}$ of about \(10^{35}\)--\(10^{36}\) erg. These bolometric energy values are comparable to those observed in stellar superflares (\(\geq 10^{33}\) erg). The specific values of flare energies are summarized in Table~\ref{tab2}.
\begin{figure*}
\centering
\includegraphics[width=10.5cm,height=8.5cm]{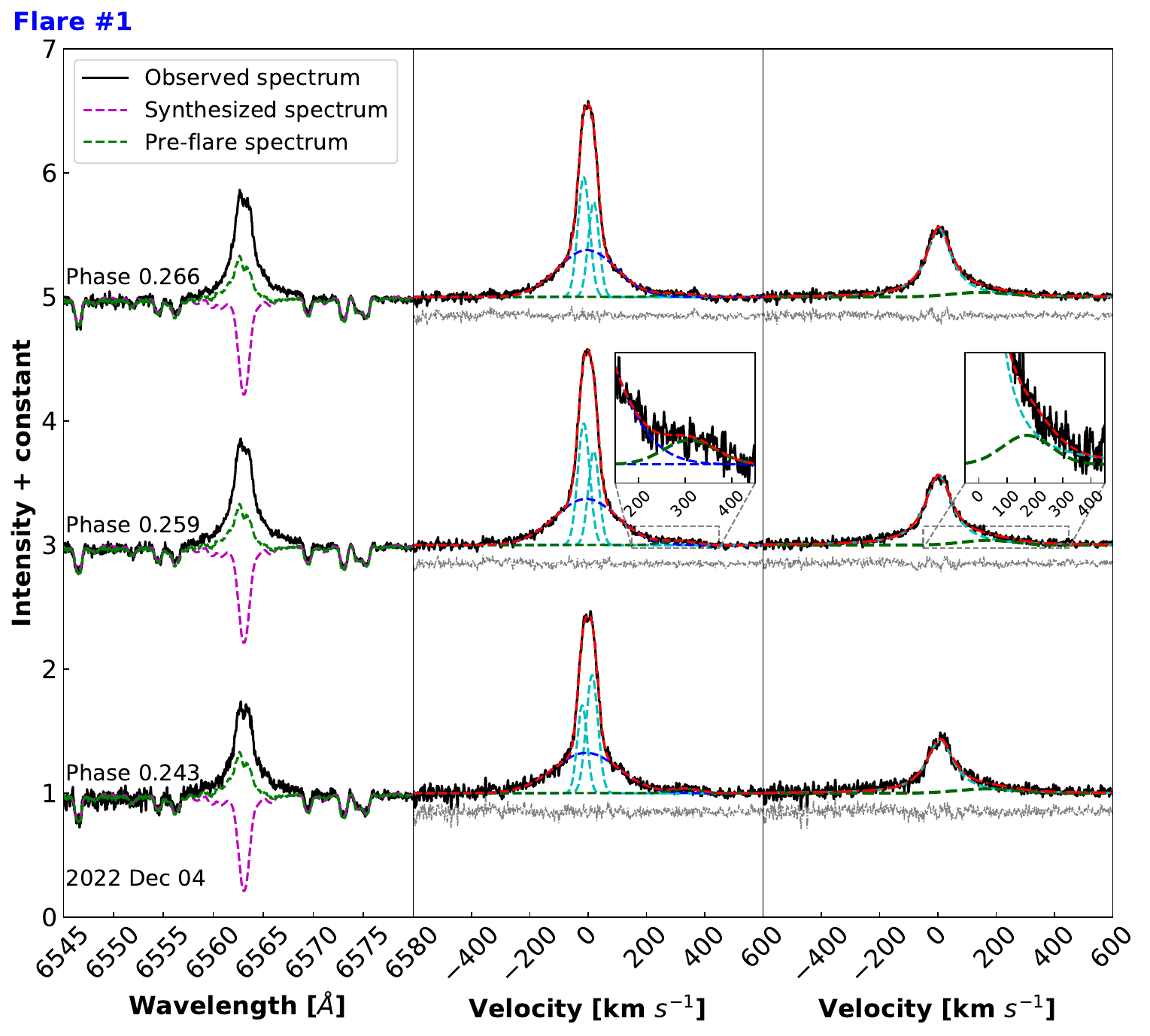}
\includegraphics[width=7.25cm,height=5.25cm]{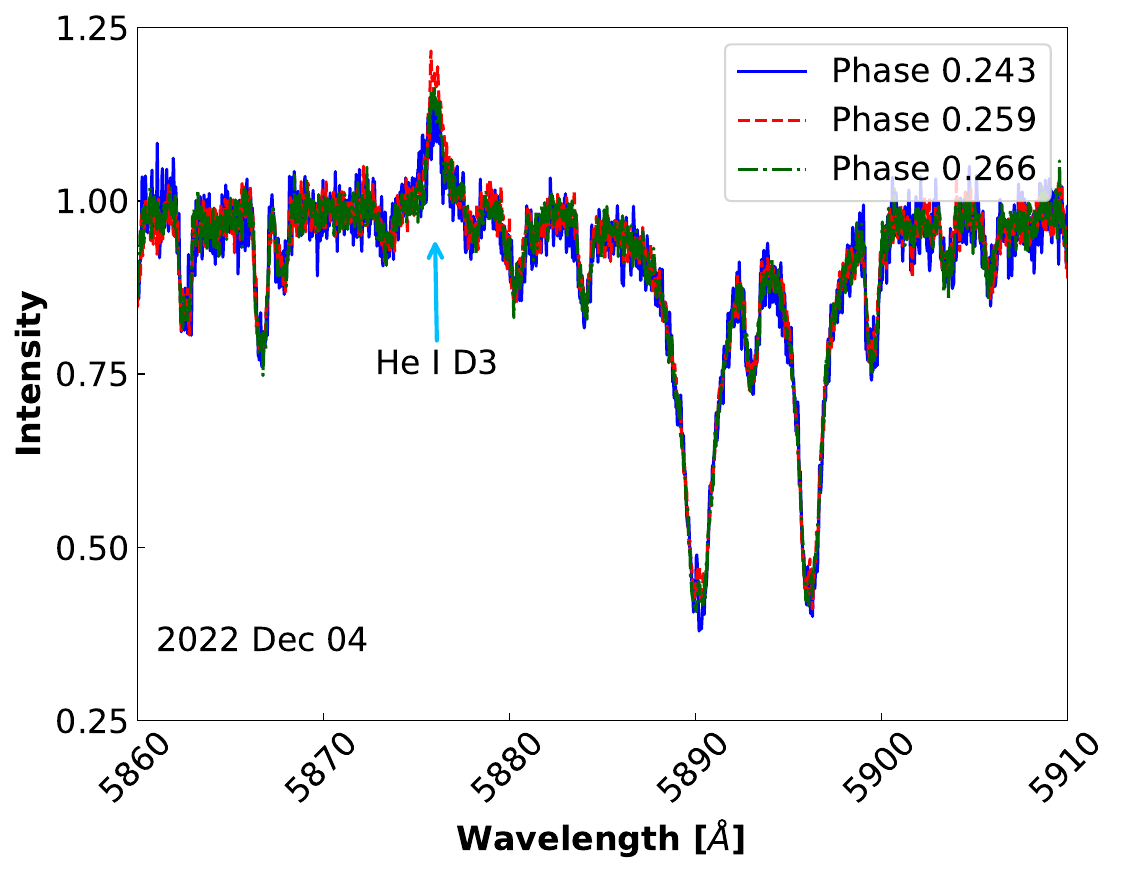}
\caption{Left panel: The H$_{\alpha}$ line profiles obtained on 2022~December~4. The observed spectra (black solid lines), the synthesized spectra (magenta dashed lines), and the preflare reference spectra (green dashed lines) are plotted in the left column. The middle column presents the STARMOD subtracted spectra (solid black lines) alongside their corresponding fittings (colored dashed lines), as well as the residuals (gray dash-dotted lines) between them. In the right column, the differential spectra (black solid lines) obtained by subtracting the preflare reference spectrum from the observed flaring spectra are presented, along with their corresponding fittings (colored dashed lines). The residuals (gray dash-dotted lines) between the differential spectra and their respective fittings are also shown below the spectra. All spectra have been corrected to the rest velocity frame of the K2~IV primary star of II~Peg. The insets show zooms into the spectral ranges with the Doppler-shifted emission signature. Right panel: The $\mbox{He~{\sc i}}$ D$_{3}$ line profiles observed on 2022~December~04. The orbital phases and observing date are also marked in the figure.}
\label{Fig1}
\end{figure*}
\begin{figure*}
\centering
\includegraphics[width=10.5cm,height=8.5cm]{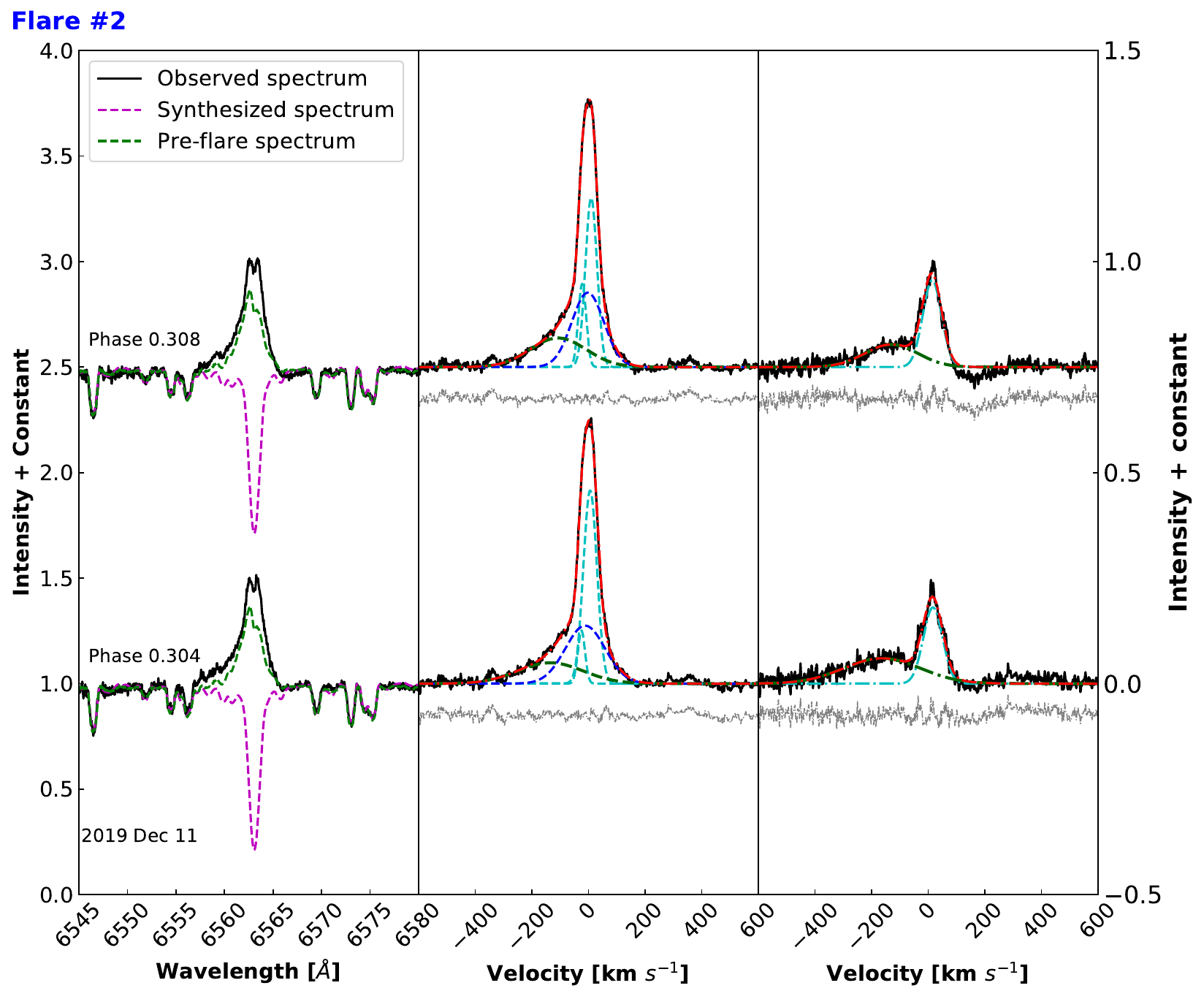}
\includegraphics[width=7.25cm,height=5.25cm]{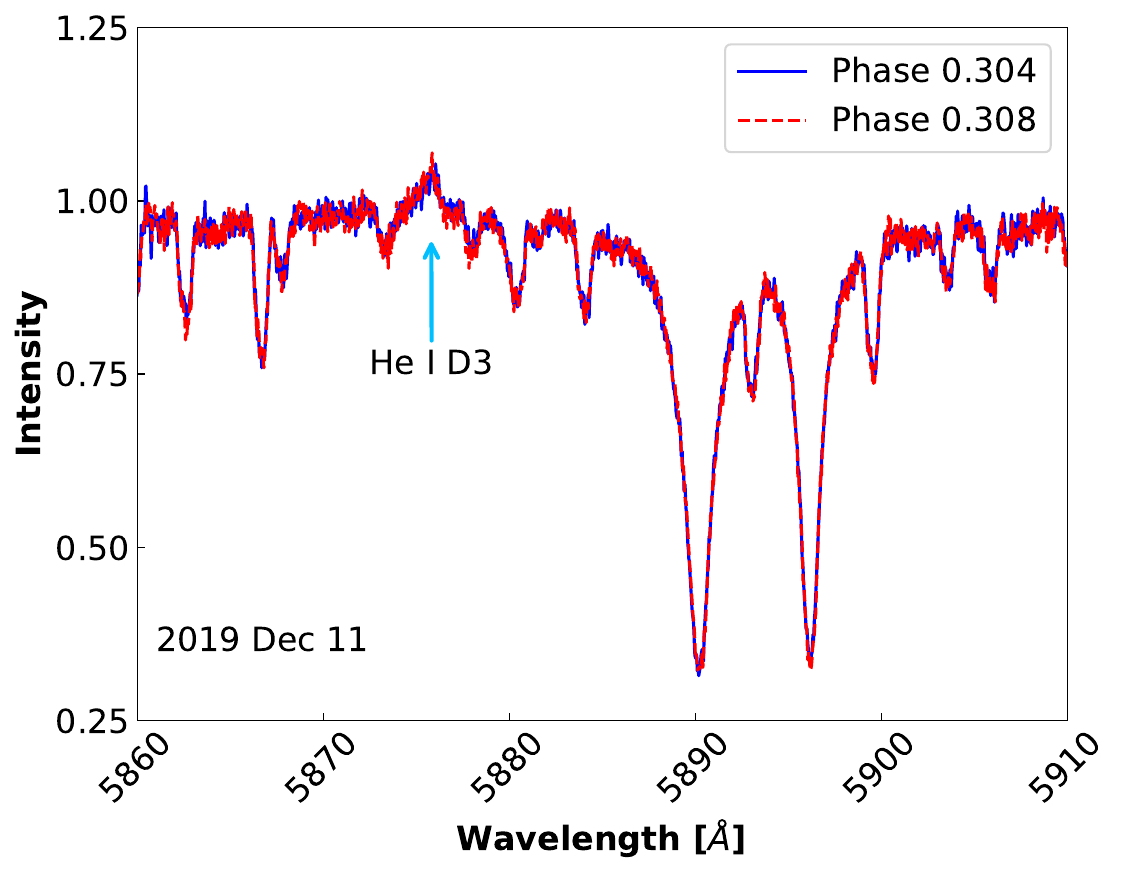}
\caption{Same as Figure~\ref{Fig1}, but for the spectra obtained on 2019~December~11. The right column of the left panel utilizes the scale of the right axis.}
\label{Fig201912}
\end{figure*}

\section{Doppler-shifted extra emission signatures during the flare events}\label{sec4}
There are five cases in which the STARMOD subtracted line profiles exhibit potential asymmetric signatures during flare events, particularly in the H$_{\alpha}$ line profile. For these observations, further analyses are performed based on the differential spectra. These additional analyses serve dual purposes: firstly, they validate the presence of asymmetric features in the line profiles; secondly, they confirm that these asymmetric features are indeed associated with the flare events. In Table~\ref{tab3}, we list the observing information, the mean signal to noise ratio (S/N) of the H$_{\alpha}$ line profile, and the measurement of the Doppler-shifted emission signature identified in the differential spectrum with maximum EW, including the bulk velocity (V$_{bulk}$), the maximum velocity (V$_{max}$), and the EW. The bulk velocity is the centroid velocity of the fitting for the Doppler-shifted emission signatures, while the maximum velocity is obtained by using V$_{max}$~=~V$_{bulk}$~$\pm$~2$\sigma$. Detailed analyses are presented as follows.

\subsection{Case on 2022~December~4}
Three spectra were observed on 2022~December~4 at phases~0.243, 0.259, and 0.266, respectively. The H$_{\alpha}$ line profiles are shown in Figure~\ref{Fig1}, which includes the STARMOD subtractions, the differential spectra, and their corresponding fittings. In Figure~\ref{Fig1}, the observed $\mbox{He~{\sc i}}$ D$_{3}$ line profiles are also displayed. It is evident that obvious emission features appeared in the $\mbox{He~{\sc i}}$ D$_{3}$ line region, indicating that a strong optical flare event happened during the observation. The emission feature in the second spectrum is much stronger than the emission in the other two spectra. 

It can be seen from the figure that the H$_{\alpha}$ subtractions can be accurately modeled using several Gaussian components, primarily consisting of two narrow emission components and one broad emission component. This is reasonable, because the H$_{\alpha}$ subtraction includes the intrinsic chromospheric emission and flare radiation, and the distribution of chromospheric activity regions is not uniformly represented over the stellar surface. Notably, a far redshifted emission component appeared in the fitting of the H$_{\alpha}$ subtraction at phase~0.259. Similar emission components could also be found in the other two spectra, but much weaker. This indicates that the H$_{\alpha}$ line profiles exhibit asymmetric features during the flare. However, in the case of the H$_{\beta}$ line subtractions, no similar far redshifted extra emission components could be detected probably due to its low S/N.

For the H$_{\alpha}$ differential spectra, it is evident that the profiles exhibit Lorentzian characteristics. In general, the flaring H$_{\alpha}$ spectra can be well fitted with Lorentzian-like profiles due to the strong Stark broadening \citep{Kowalski2017, Namekata2020}. Consequently, we employ a Lorentzian function to model the symmetric features of these profiles, in combination with a Gaussian function to account for any potential asymmetric features. A more redshifted emission component is needed to fit the differential spectra, particularly in the second spectrum at phase~0.259, which exhibits a more pronounced extra emission component with a bulk velocity of about 148~km~s$^{-1}$. Additionally, similar to the results of the STARMOD subtractions, there are no comparable redshifted emission components observed in the H$_{\beta}$ differential spectra. 

The EWs of the H$_{\alpha}$ subtractions are 3.845~$\pm$~0.046~\AA~at phase~0.243, 4.213~$\pm$~0.045~\AA~at phase~0.259, and 4.060~$\pm$~0.047~\AA~at phase~0.266, while the EWs of the H$_{\beta}$ subtractions are 1.692~$\pm$~0.030~\AA, 2.060~$\pm$~0.008~\AA, and 1.868~$\pm$~0.015~\AA, respectively. Consistent variations are also observed in the EWs of the H$_{\alpha}$ and H$_{\beta}$ differential spectra (see Table~\ref{tab2}). The EW variatios suggest a rapid emission increase from phase 0.243 to 0.259 and then a gradual emission decrease to phase 0.266, which is consistent with the changes of the $\mbox{He~{\sc i}}$~D$_{3}$ emission.

\subsection{Case on 2019~December~11}
During the observation of 2019~December~11, two spectra were observed at phases~0.304 and 0.308, respectively. Figure~\ref{Fig201912} shows the H$_{\alpha}$ line profiles, their STARMOD subtractions, the differential spectra, as well as their corresponding fittings. Additionally, Figure~\ref{Fig201912} presents the profiles of the $\mbox{He~{\sc i}}$ D$_{3}$ line. It is evident that a significant emission feature appeared in the $\mbox{He~{\sc i}}$ D$_{3}$ line region, indicating that a strong optical flare event occurred during the observation.

The H$_{\alpha}$ subtractions can be accurately modeled using multiple Gaussian components. The fittings reveal a more blueshifted emission component, in addition to two narrow emission components and one broad emission component. This indicates that the H$_{\alpha}$ line profiles exhibit the blueshifted asymmetric features during the flare. Furthermore, similar blueshifted emission components are also seen in the H$_{\beta}$ subtractions. However, due to the significantly low S/N of the H$_{\beta}$ line, we do not perform any further analysis and do not present its profiles here.

For the H$_{\alpha}$ differential spectra, the line profiles exhibit Gaussian characteristics rather than Lorentzian ones. The application of Gaussian fitting effectively models these differential profiles. In addition to the central narrow emission component, it is evident that a more blueshifted emission component exists in their spectra. This component is characterized by a bulk velocity of approximately -163~km~s$^{-1}$ at phase 0.304.

The EWs of the H$_{\alpha}$ subtractions are 3.057~$\pm$~0.016~\AA~at phase~0.304 and 2.914~$\pm$~0.016~\AA~at phase~0.308, respectively, while the EWs of the H$_{\beta}$ subtractions are 1.195~$\pm$~0.012~\AA~and 1.1017~$\pm$~0.010~\AA, respectively. This decreased EW variation is also evident in both of the H$_{\alpha}$ and H$_{\beta}$ differential spectra. The decline in EW variations for both H$_{\alpha}$ and H$_{\beta}$ lines suggests that our observations likely captured a gradual decay of the flare event.
\begin{figure*}
\centering
\includegraphics[width=10.5cm,height=8.5cm]{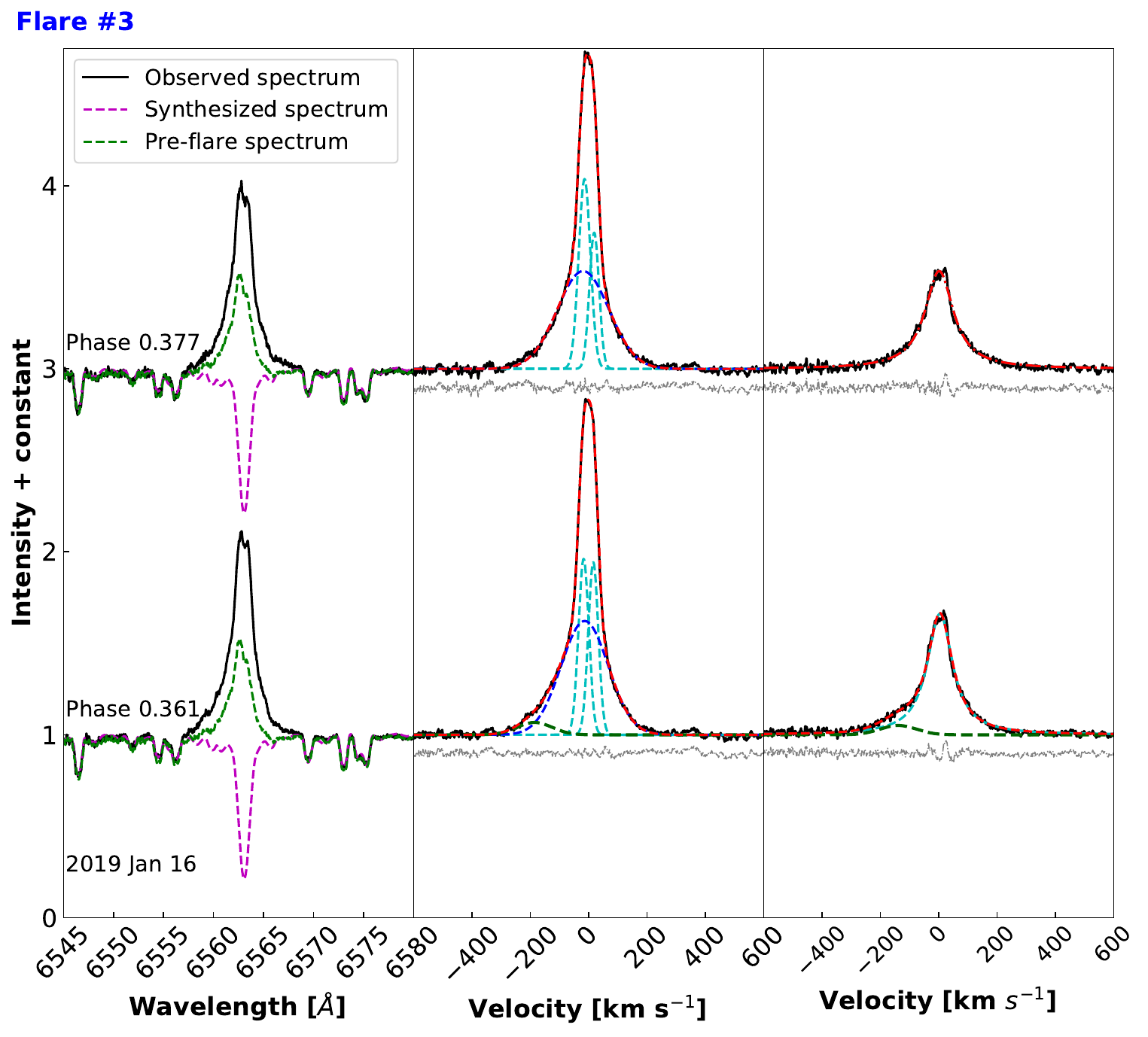}
\includegraphics[width=7.25cm,height=5.25cm]{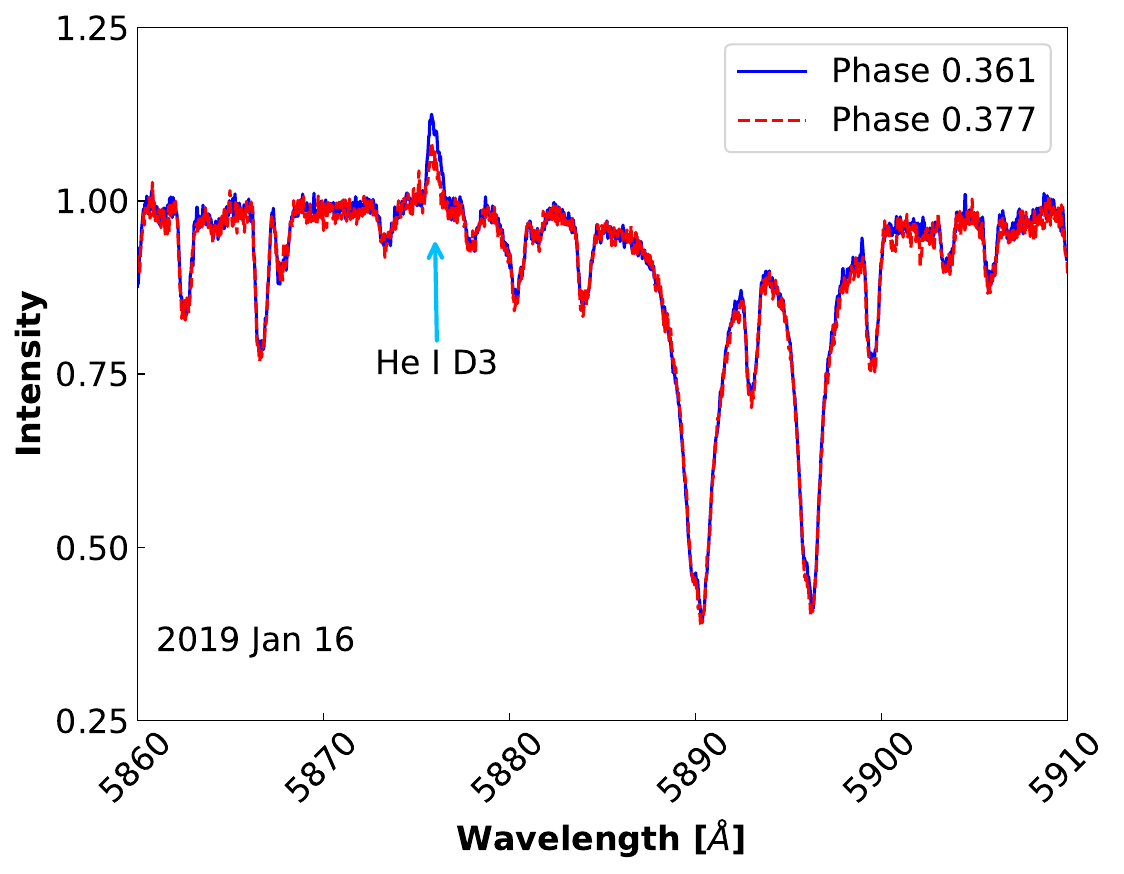}
\caption{Same as Figure~\ref{Fig1}, but for the spectra obtained on 2019~January~16. }
\label{Fig2}
\end{figure*}
\begin{figure*}
\centering
\includegraphics[width=8.9cm,height=6.5cm]{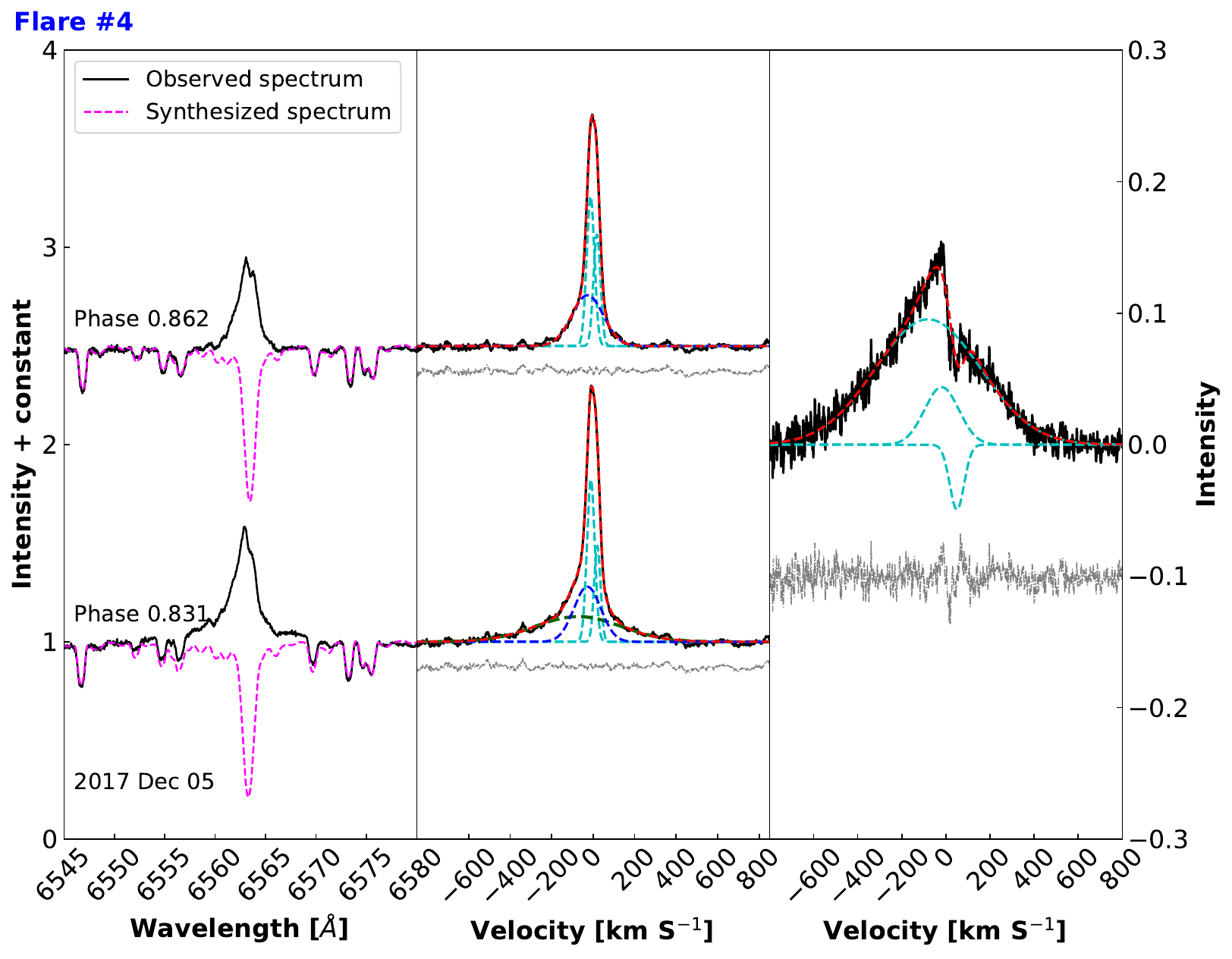}
\includegraphics[width=8.9cm,height=6.5cm]{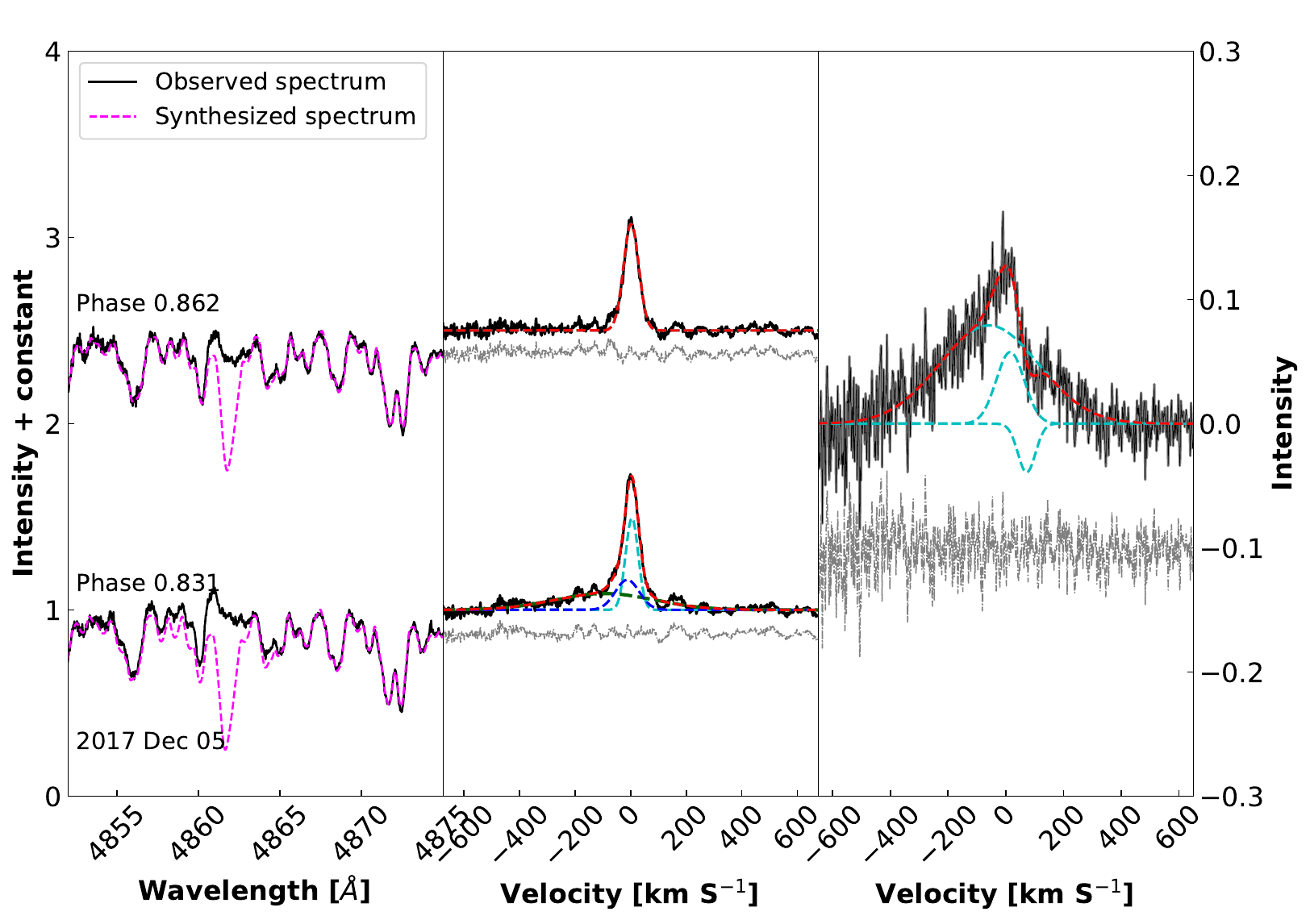}
\includegraphics[width=8.9cm,height=4.5cm]{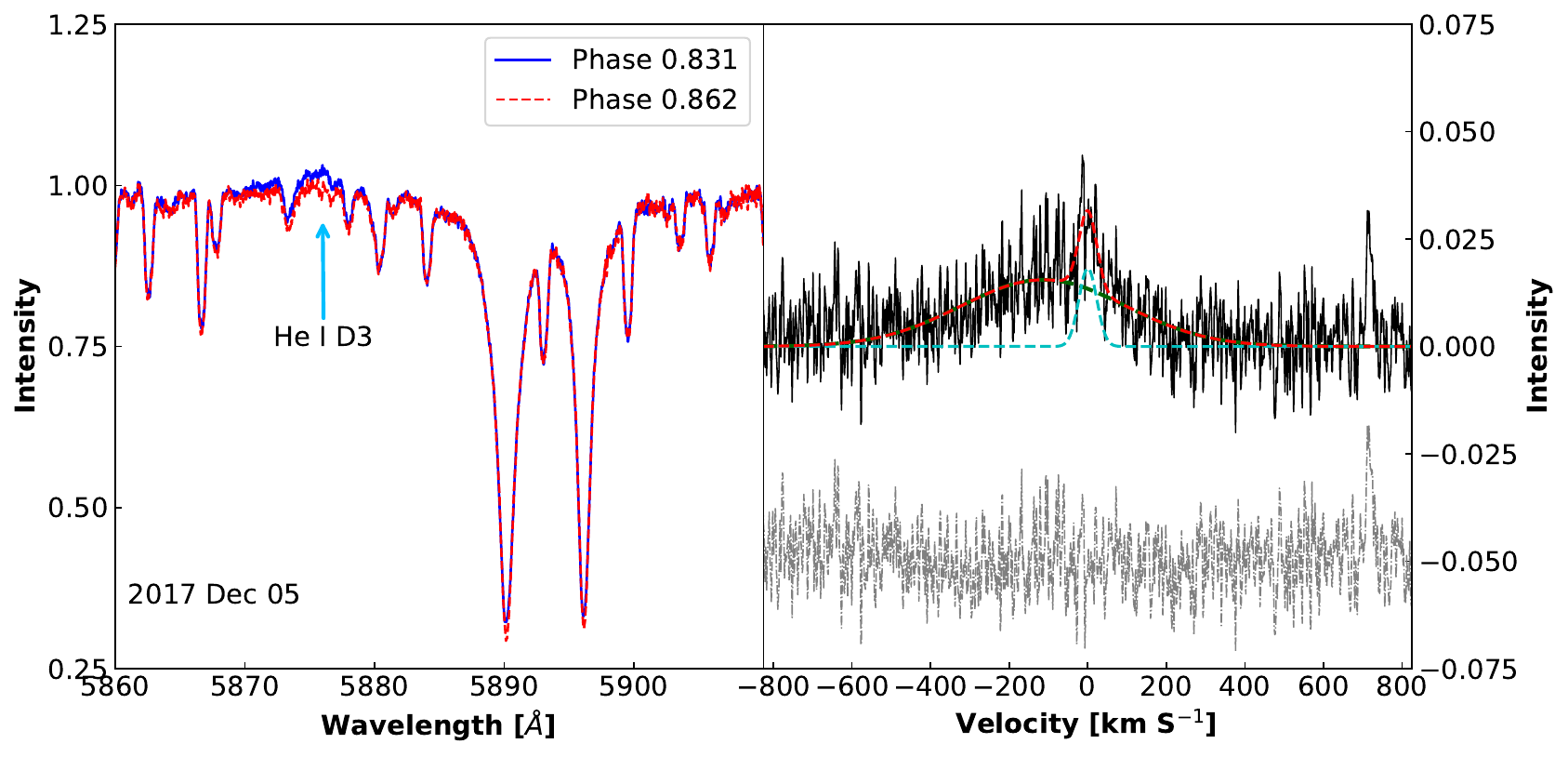}
\includegraphics[width=8.9cm,height=4.5cm]{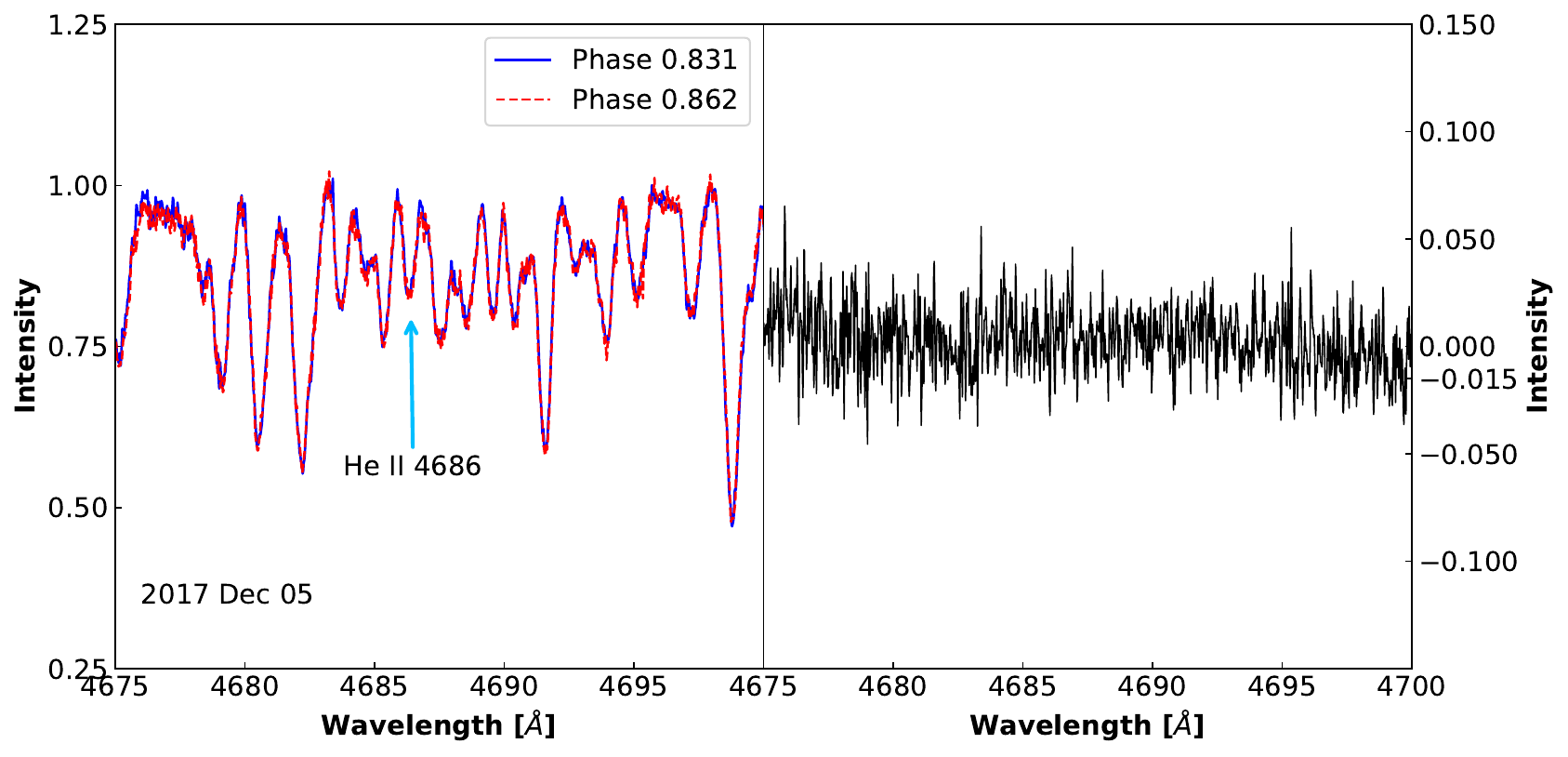}
\caption{Upper panels: Same as Figure~\ref{Fig1}, but for the H$_{\alpha}$ and H$_{\beta}$ spectra obtained on 2017~December~05. Lower panels: The $\mbox{He~{\sc i}}$ D$_{3}$ (left) and $\mbox{He~{\sc ii}}$~4686~\AA~(right) line profiles obtained on 2017~December~05. The differential spectra (black solid line) between two spectra are plotted in the right sections of both panels. For the $\mbox{He~{\sc i}}$ D$_{3}$ line, we model the differential spectrum by using Gaussian fitting (dashed lines). All spectra have been corrected to the rest velocity frame of the K2~IV primary star of II~Peg. The right section of each panel employs the scale of the right axis.}
\label{Fig3}
\end{figure*}

\subsection{Case on 2019~January~16}
Two spectra were observed on 2019~January~16 at phases~0.361 and 0.377, respectively. The H$_{\alpha}$ line profiles are illustrated in Figure~\ref{Fig2}, which includes the STARMOD subtractions, the differential spectra, and their corresponding fittings. Additionally, Figure~\ref{Fig2} presents the $\mbox{He~{\sc i}}$ D$_{3}$ line profiles. It is evident that a significant emission feature appeared in the $\mbox{He~{\sc i}}$ D$_{3}$ line region, indicating that an optical flare occurred. The emission characteristic in the first spectrum observed at phase 0.361 is more pronounced than that in the second spectrum.

The H$_{\alpha}$ subtractions can be accurately modeled using multiple Gaussian components, primarily consisting of two narrow emission components and one broad emission component. Notably, a far blueshifted emission component appears in the fitting of the H$_{\alpha}$ subtraction at phase~0.361. This observation indicates that the H$_{\alpha}$ line profile show asymmetric feature. Furthermore, no analogous emission feature is observed in the H$_{\beta}$ subtractions, probably due to the low S/N. 

For the H$_{\alpha}$ differential spectra, similar to the analysis conducted on 2022 December~4, we model the profiles using a Lorentzian function for the symmetric feature and a Gaussian function for the asymmetric feature. It is evident that a blueshifted excess emission component exists in the first spectrum, characterized by a bulk velocity of about -139~km~s$^{-1}$. Furthermore, similar to the result of the STARMOD subtraction, there is no comparable emission component appeared in the H$_{\beta}$ differential spectra.

The EWs of the H$_{\alpha}$ subtractions are 4.828~$\pm$~0.043~\AA~at phase~0.361 and 4.453~$\pm$~0.053~\AA~at phase~0.377, respectively, while the EWs of the H$_{\beta}$ subtractions are recorded as 1.723~$\pm$~0.012~\AA~and 1.444~$\pm$~0.022~\AA, respectively. The observed decrease in EW variation, which is also apparent in both the H$_{\alpha}$ and H$_{\beta}$ differential spectra (see Table \ref{tab2}), suggests that the observation likely captured a gradual decay of the flare event. This is also consistent with the change of the $\mbox{He~{\sc i}}$ D$_{3}$ line emission.

\subsection{Case on 2017~December~5}
During the observation of 2017~December~5, two spectra were obtained at phase~0.831 and 0.862, respectively. The H$_{\alpha}$ and H$_{\beta}$ line profiles are presented in Figure~\ref{Fig3}. It is evident that the line profiles observed at phase~0.831 exhibit broader and more pronounced wings compared to those at phase~0.862. Additionally, we present the $\mbox{He~{\sc i}}$~D$_{3}$ line profiles in Figure~\ref{Fig3}. Notably, the first spectrum obtained at phase 0.831 reveals a significantly broader emission feature when contrasted with the second spectrum. These observing characteristics suggest that the first spectrum is indicative of a flaring event, whereas the second spectrum does not correspond to a flaring state.

For the H$_{\alpha}$ and H$_{\beta}$ subtractions, as shown in Figure~\ref{Fig3}, we model the profiles using the Gaussian fitting method. In addition to a narrow emission component and a broad emission one, it is noteworthy that a more broad and blueshifted emission component can be found in the fitting of the first subtracted H$_{\alpha}$ spectrum. Moreover, the H$_{\beta}$ line profile shows the similar behavior that a broad and blueshifted emission component appears in the fitting of the first subtraction.

Unlike using the mean spectrum as the preflare reference spectrum, here we employ the second spectrum as the preflare reference for this flare event. The differential spectra of both the H$_{\alpha}$ and H$_{\beta}$ lines between the first and second spectra are also presented in Figure~\ref{Fig3}. We model these differential spectra using Gaussian fitting method. It can be seen that the blueshifted broad emission exists, as the results of the STARMOD subtractions. Furthermore, to accurately fit the profiles of the differential H$_{\alpha}$ and H$_{\beta}$ lines, two redshifted absorption profiles are incorporated into the fitting process. This phenomenon may be attributed to the presence of downward moving material during the flare.

For the $\mbox{He~{\sc i}}$~D$_{3}$ line, the much broader emission feature observed at phase 0.831 not only indicates the occurrence of a significant optical flare but also suggests that there may be a common emission source responsible for both the broad and blueshifted emission components of the H$_{\alpha}$ and H$_{\beta}$ lines. Using the same analytical method applied to the H$_{\alpha}$ and H$_{\beta}$ lines, the differential spectrum of the $\mbox{He~{\sc i}}$ D$_{3}$ line between two spectra is illustrated in Figure~\ref{Fig3}. This differential profile can well be fitted by a narrow emission component and a blueshifted broad emission. Furthermore, as illustrated in Figure~\ref{Fig3}, the $\mbox{He~{\sc ii}}$~4686~\AA line is also included in our spectral analysis. However, no corresponding emission feature is detected in the differential spectrum of the $\mbox{He~{\sc ii}}$~4686~\AA line.

During our observation, the detection of only a single flaring spectrum makes it challenging to determine which stage the flare is in.

\begin{figure*}
\centering
\includegraphics[width=10.5cm,height=8.5cm]{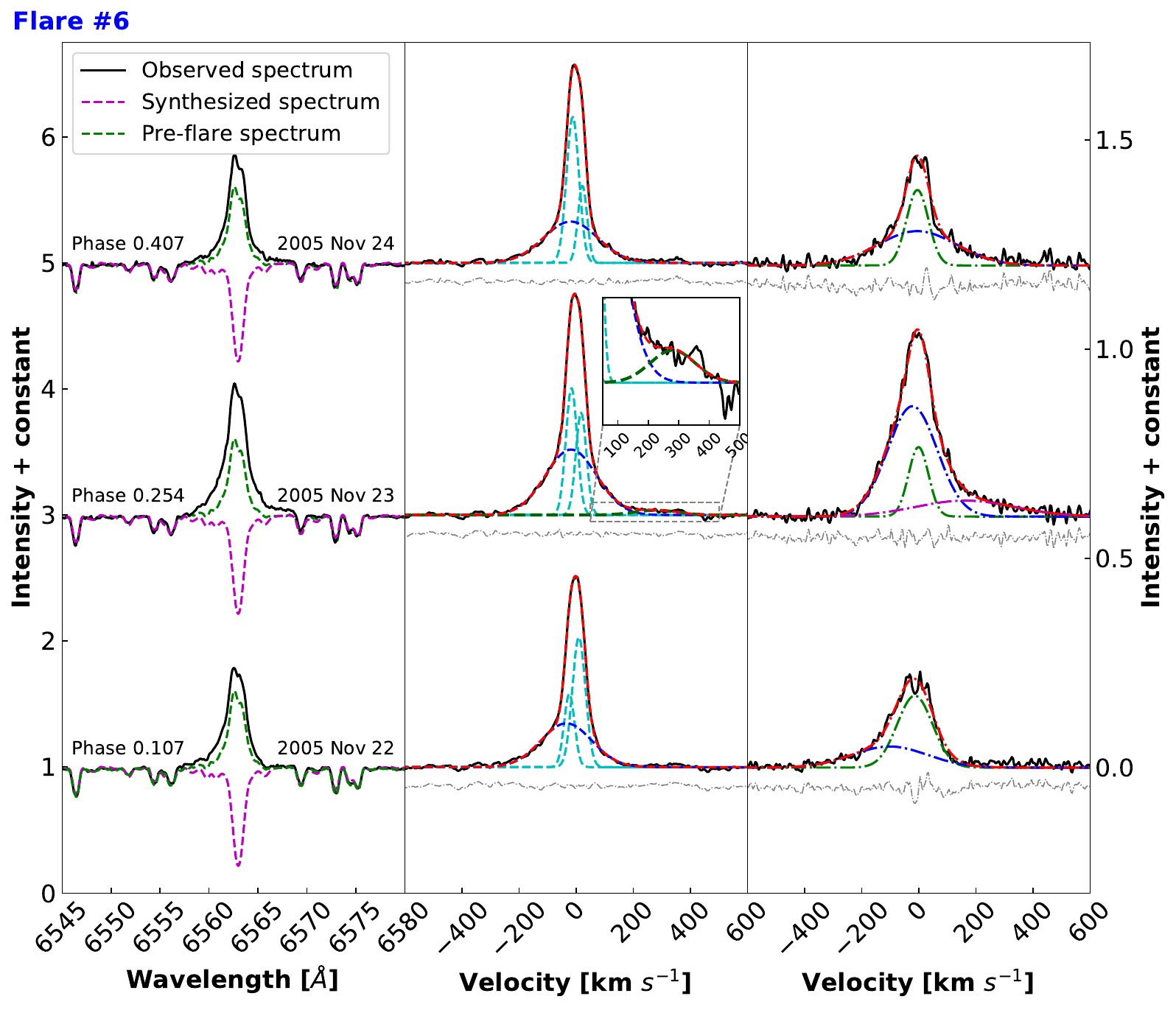}
\includegraphics[width=7.25cm,height=5.25cm]{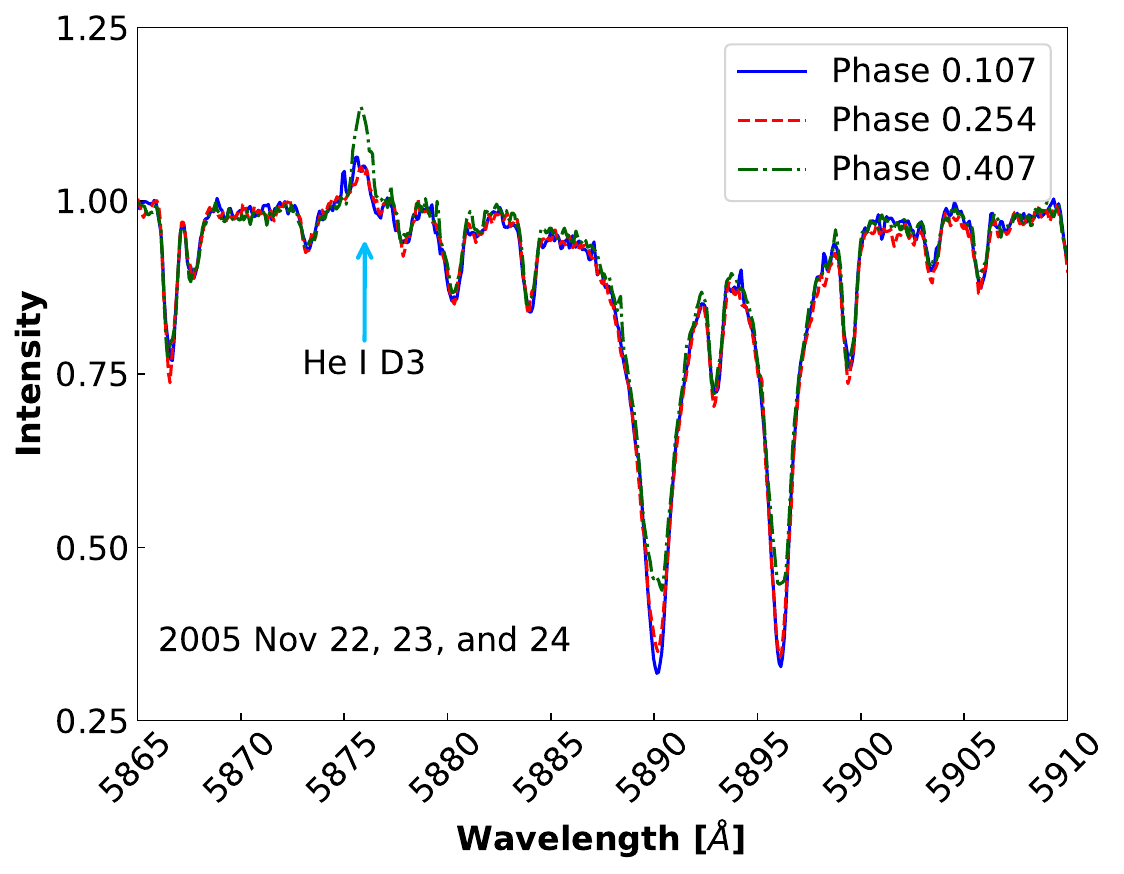}
\caption{Same as Figure~\ref{Fig1}, but just for the H$_{\alpha}$ and $\mbox{He~{\sc i}}$~D$_{3}$ spectra observed on 2005~November~22, 23, 24. The right column of the left panel utilizes the scale of the right axis.}
\label{Fig4}
\end{figure*}
\begin{deluxetable*}{cccccccccc}
\tablenum{3}
\tablecaption{Parameters of the CME Candidates\label{tab3}}
\tablewidth{0pt}
\tablehead{
\colhead{\bf{No}}&\colhead{\bf{UT date}}&\colhead{\bf{Phase}}&\colhead{\bf{Mean H$_{\alpha}$}}&\colhead{\bf{Enhanced}}& \colhead{\bf{V$_{bulk}$}}&\colhead{\bf{V$_{max}$}}&\colhead{\bf{EW}}&\colhead{\bf{M$_{CME}$}}&\colhead{\bf{E$_{kin}$}}\\
\nocolhead{}&\nocolhead{}&\nocolhead{}&\colhead{\bf{S/N}}&\colhead{\bf{wing}}&\colhead{\bf{(km s$^{-1}$)}}&\colhead{ \bf{(km s$^{-1}$)}}&\colhead{\bf{(\AA)}}&\colhead{\bf{(g)}}&\colhead{\bf{(erg)}}
}
\startdata
1 & 2022~Dec~4  & 0.259 & 62  & Red  & 148~$\pm$~19  & 350  & 0.191~$\pm$~0.009 & $6.9~\times~10^{19}$ & $7.6~\times~10^{33}$\\
2 & 2019~Dec~11  & 0.304 & 96  & Blue & -163~$\pm$~5  & -434 & 0.425~$\pm$~0.005 & $1.5~\times~10^{20}$ & $2.0~\times~10^{34}$\\
3 & 2019~Jan~16  & 0.361 & 116 & Blue & -139~$\pm$~4  & -264 & 0.173~$\pm$~0.003 & $6.2~\times~10^{19}$ & $6.0~\times~10^{33}$\\
4 & 2017~Dec~5  & 0.831 & 191 & Blue & -81~$\pm$~2   & -560 & 1.219~$\pm$~0.003 & $4.4~\times~10^{20}$ & $1.4~\times~10^{34}$\\
5 & 2005~Nov~23  & 0.254 & 212 & Red  & 162~$\pm$~45  & 507  & 0.345~$\pm$~0.010 & $1.2~\times~10^{20}$ & $1.6~\times~10^{34}$\\
6 & 2005~Nov~22  & 0.107 & 192 & Blue & -102~$\pm$~13 & -365 & 0.364~$\pm$~0.017 & $1.3~\times~10^{20}$ & $6.8~\times~10^{33}$\\
\enddata
\end{deluxetable*}
\begin{figure*}
\centering
\includegraphics[width=6.9cm,height=13cm]{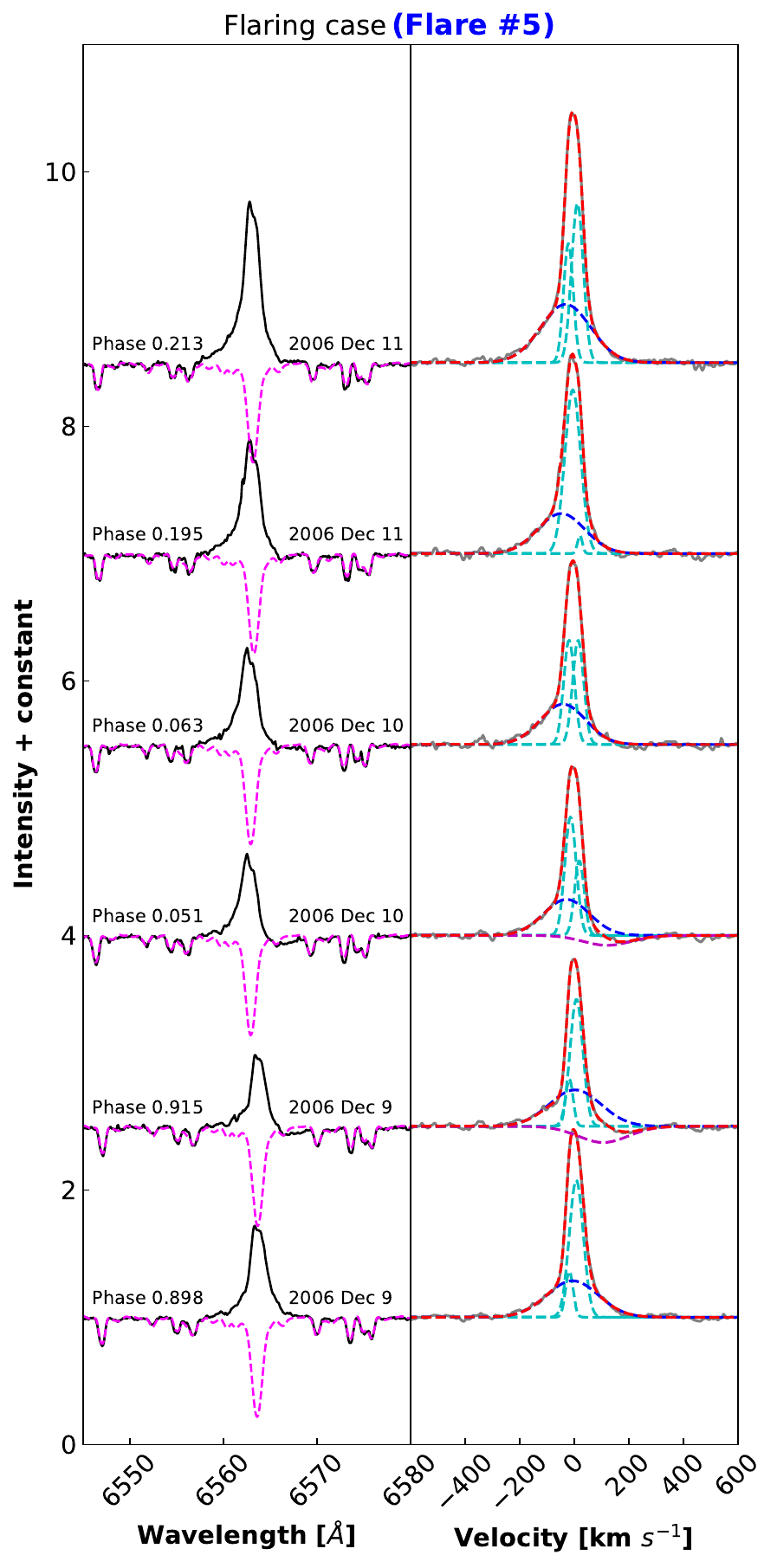}
\includegraphics[width=6.9cm,height=13cm]{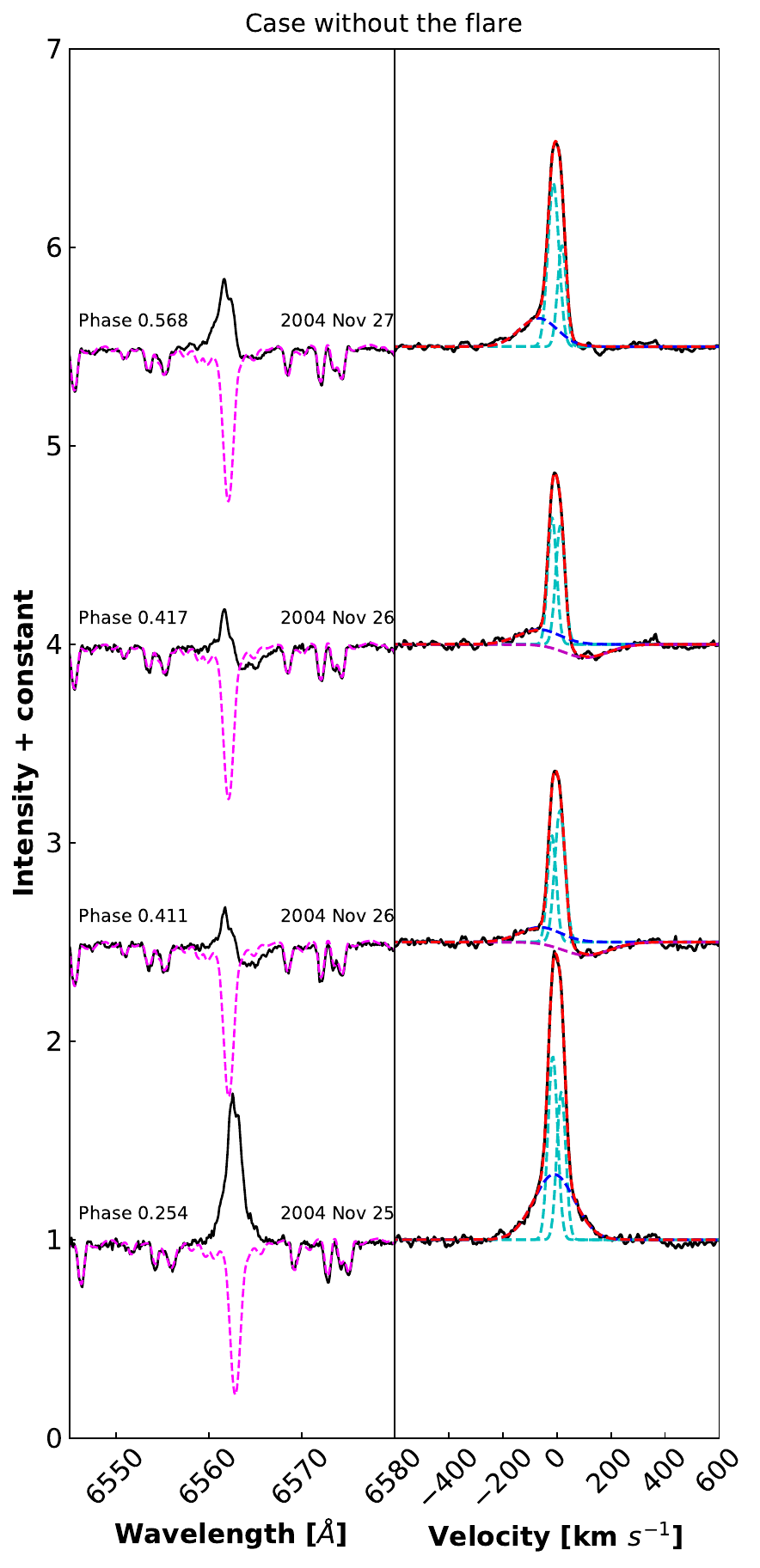}
\caption{Same as Figure~\ref{Fig1}, but just for the H$_{\alpha}$ spectra observed from 2006~December~9 to 11 and from 2004~November~25 to 27, respectively.}
\label{Fig11}
\end{figure*}

\subsection{Case on 2005~November~22 to 24}
During consecutive observing nights from 2005~November~22 to 24, we obtained three spectra at phases of 0.107, 0.254, and 0.407. The H$_{\alpha}$ line profiles are illustrated in Figure~\ref{Fig4}, which also includes the STARMOD subtractions, the differential spectra, and the corresponding fittings. Additionally, the $\mbox{He~{\sc i}}$~D$_{3}$ profiles are presented in Figure~\ref{Fig4}. The $\mbox{He~{\sc i}}$~D$_{3}$ line shows emission features above the continuum during our observations, which means that either a long-duration optical flare or different flares occurred during the three nights. Due to the poor temporal resolution, it is difficult to identify them.

The H$_{\alpha}$ subtractions can be well modeled using several Gaussian components, primarily consisting of two narrow emission components and one broad emission component. For the first spectrum observed at phase~0.107 on 2005~November~22, the fitted broad emission component is more blueshifted, indicating that the H$_{\alpha}$ profile displays a blueshifted asymmetric feature. Moreover, for the second spectrum obtained at phase~0.254 on 2005~November~23, a far redshifted emission component is shown in the subtraction, further suggesting the presence of a redshifted asymmetric feature during the flare.

For the H$_{\alpha}$ differential spectra, we employ Gaussian fitting method to model the profiles. It is evident that there exists a significantly redshifted emission component with a bulk velocity of about 162~km~s$^{-1}$ in the second spectrum. Additionally, a blueshifted emission component was detected in the first differential spectrum, exhibiting a bulk velocity of about -102~km~s$^{-1}$.

The EWs of the H$_{\alpha}$ subtractions are 3.629~$\pm$~0.103~\AA~at phase 0.107, 4.635~$\pm$~0.034~\AA~at phase 0.254, and 3.689~$\pm$~0.093~\AA~at phase 0.407. The EWs exhibit an increasing trend from phase 0.107 to 0.254, followed by a decreasing trend to phase~0.407. The similar variation can also be observed in the H$_{\alpha}$ differential spectra (see Table~\ref{tab2}). The $\mbox{He~{\sc i}}$ D$_{3}$ line emission in the third spectrum is significantly stronger than the ones in the previous two spectra, which contradicts the EW variations of the H$_{\alpha}$ subtractions and differential spectra. Therefore, one possible explanation is the occurrence of different flares. In Section~\ref{sec3}, we assume that the stronger emission in these three spectra originates from a long-duration flare event, and the flare energy has been derived.

\begin{deluxetable*}{cccccc}
\tablenum{4}
\tablecaption{Measurements of the Redshifted Excess Absorption Signatures\label{tab4}}
\tablewidth{0pt}
\tablehead{
\colhead{\bf{No}}&\colhead{\bf{UT date}}&\colhead{\bf{Phase}}&\colhead{\bf{Accompanied by}}&\colhead{\bf{$V_{bulk}$}}&\colhead{\bf{EW}}\\
\nocolhead{}&\nocolhead{}&\nocolhead{}&\colhead{\bf{flare}}&\colhead{\bf{(km s$^{-1}$)}}&\colhead{\bf{(\AA)}}
}
\startdata
Absorption~1& 2006~Dec~09 & 0.915 & Yes & 106~$\pm$~36 & -0.629~$\pm$~0.010\\
& 2006~Dec~10 & 0.051 & Yes & 120~$\pm$~78 & -0.373~$\pm$~0.013\\
\hline
Absorption~2& 1999~July~27 & 0.581 & Yes & 143~$\pm$~3   & -0.087~$\pm$~0.002\\
& 1999~July~27 & 0.584 & Yes & 142~$\pm$~4   & -0.088~$\pm$~0.002\\
& 1999~July~27 & 0.588 & Yes & 140~$\pm$~4   & -0.103~$\pm$~0.002\\
& 1999~July~27 & 0.591 & Yes & 140~$\pm$~3   & -0.111~$\pm$~0.002\\
& 1999~July~27 & 0.594 & Yes & 141~$\pm$~3   & -0.096~$\pm$~0.001\\
& 1999~July~27 & 0.597 & Yes & 145~$\pm$~3   & -0.101~$\pm$~0.003\\
\hline
Absorption~3& 2019~Dec~10 & 0.146 & No & 87~$\pm$~112 & -0.288~$\pm$~0.010\\
& 2019~Dec~10 & 0.158 & No & 76~$\pm$~196 & -0.549~$\pm$~0.011\\
& 2019~Dec~10 & 0.162 & No & 82~$\pm$~124 & -0.570~$\pm$~0.013\\
\hline
Absorption~4& 2004~Nov~26 & 0.411 & No & 109~$\pm$~33 & -0.309~$\pm$~0.006\\
& 2004~Nov~26 & 0.417 & No & 108~$\pm$~24 & -0.288~$\pm$~0.010\\
\hline
Absorption~5& 1999~Dec~22 & 0.552 & No & 97~$\pm$~29   & -0.241~$\pm$~0.006 \\
& 1999~Dec~22 & 0.556 & No & 86~$\pm$~23   & -0.251~$\pm$~0.006 \\
\enddata
\end{deluxetable*}

\section{Redshifted extra absorption signatures}\label{sec5}
Besides the Doppler-shifted emission signatures, there are some extra absorptions in the subtracted H$_{\alpha}$ line profiles. The extra absorption signatures always appear redshifted with respect to the line centre, which could be found not only in the flaring spectra but also in the spectra when no flare happened. 

\subsection{Cases during the flare events}
An example of the redshifted extra absorption signatures during the flares can be found in Figure~\ref{Fig11}, which shows the H$_{\alpha}$ spectra obtained from 2006 December~9 to 11, together with their subtractions. From the figure, it can be seen that there are extra absorption signatures in the red wings of the second H$_{\alpha}$ subtraction on 2006 December~9 (at phase 0.915) and the first subtraction on 2006 December~10 (at phase 0.051). The corresponding $\mbox{He~{\sc i}}$ D$_{3}$ lines obtained simultaneously show obvious emission features, which suggests that a long duration optical flare or different flare events happened during these three consecutive days. Therefore, the appearance of these transient absorption signatures is probably associated with the flare(s). Similar flare-related redshifted extra absorption signatures can also be found in the H$_{\alpha}$ subtractions on 1999 July~27. With the H$_{\alpha}$ subtractions, as shown in Figure~\ref{Fig11}, we model the profiles by using the Gaussian fitting method, in which one absorption Gaussian profile is used to represent the excess absorption feature. The measurements of the redshifted extra absorption features, including the observing information, the bulk velocities, and the EWs, are listed in Table~\ref{tab4}.

\subsection{Cases without flares}
Figure~\ref{Fig11} also shows the H$_{\alpha}$ spectra obtained from 2004 November~25 to 27, together with their subtractions. It can be seen that there are extra absorption signatures in the red wings of the H$_{\alpha}$ subtractions on 2004 November~26. Unlike the situation described above, there is no emission feature around the corresponding $\mbox{He~{\sc i}}$ D$_{3}$ line region, which suggests that there is no optical flare event happened during the observation. Therefore, it may imply that the appearance of the extra absorption signatures is not always associated with flares. Moreover, there are similar extra absorptions signatures in the H$_{\alpha}$ subtractions on 2019 December~10 and 1999 December~22. We model the H$_{\alpha}$ subtractions by using the Gaussian fitting method, and measurements of these extra absorption signatures are also presented in Table~\ref{tab4}.

\section{Discussion}\label{sec6}
\subsection{Doppler-shifted emission signatures during the flares}
\subsubsection{Possible interpretation of the origins}
As presented in Section~\ref{sec4}, we have detected six cases showing the Doppler-shifted emission signatures in the wings of the Balmer line profiles, especially for the H$_{\alpha}$ line, which include two redshifted emission components and four blueshifted emission ones. The blueshifted emission components could be thought to originate from prominence eruptions moving towards the observer, while the redshifted emission components could be interpreted as results of backward-directed prominence eruptions occurring near the stellar limb. Although the bulk velocities of these Doppler-shifted emission signatures are below the escape velocity of the K2 IV primary star of II~Peg ($\sim$306~km~s$^{-1}$), it is noteworthy that their maximum velocities approach or even exceed this escape velocity (see Table~\ref{tab3}). Moreover, due to projection effects, the measured velocities represent only lower limits of the actual eruption velocities. When prominence material is ejected near the stellar limb, its radial velocity is just a smaller component alone the line of sight. Consequently, these prominence eruptions are likely to develop into stellar CMEs. For the blueshifted emission component observed on 2017 December~5, we have a more detailed discussion in Section~6.1.3.

Moreover, it is important to note that the Doppler-shifted signatures in the spectral line profiles can also be caused by other forms of plasma motion during the flares. According to current knowledge of solar flares, the blueshifted emission component could also be attributed to upward plasma above the chromospheric evaporation \citep[e.g.][]{Tei2018}. The velocities of chromospheric evaporation typically reach several tens km~s$^{-1}$ in the solar case. In addition, the redshifted emission components can also be interpreted as evidence of chromospheric condensation or coronal rain along post-flare loops or falling material in a prominence eruption \citep{Koller2021, Wu2022}. Usually, coronal rain falls down to the solar surface at velocities of 30--200~km~s$^{-1}$, with a mean value of about 60--70~km~s$^{-1}$ \citep{Antolin2012, Lacatus2017}. Chromospheric condensation arising from downward moving material is observed with a typical velocity of several tens km~s$^{-1}$ \citep{Ichimoto1984}. Even if the velocities of the Doppler-shifted emission signatures observed in the spectral line profiles of II Peg (excluding the observation on 2017 December~5) are generally higher than the typical velocities associated with above plasma motions during solar flares, we cannot entirely rule out these possibilities based solely on velocity considerations. Because the energy scales of flares on II Peg are greater than that of typical solar flares, the magnitude of the Doppler-shifted phenomena may differ significantly from those observed in solar flares. For instance, \citet{Namizaki2023ApJ} reported H$_{\alpha}$ red asymmetry during a superflare on the M dwarf YZ~Canis Minoris and interpreted the asymmetric features with velocities ranging from 200 to 500 km~s$^{-1}$ as resulting from chromospheric condensation. Moreover, prior to the detection of the redshifted emission signature associated with falling material during a prominence eruption, as stated by \citet{Wu2022}, there should be a blueshifted feature representing the prominence erupting motion at the flare onset. Although we did not detect the blueshifted emission signature prior to the redshifted one observed on 2022 December~4, this absence is likely due to the limited temporal resolution of our observations. Therefore, we cannot rule out this possibility. For the redshifted emission signature observed on 2005 November~23, it is noteworthy that a blueshifted emission one was detected on 2005 November~22. 

\subsubsection{Masses and kinetic energies of the CME candidates}
From the Doppler-shifted emission signatures in the H$_{\alpha}$ line profiles, we can estimate the masses of CME candidates following the same procedure in C24. All details can be found there. The masses of all CME candidates are calculated and the values are in a range of $10^{19}$--$10^{20}$~g, which are in agreement with the results estimated by several authors on other active stars \citep[e.g.][]{Moschou2019, Inoue2023, Namekata2024}. Moreover, with the estimated masses and the corresponding Doppler-shifted velocities, the kinetic energies of CME candidates are inferred from
\begin{eqnarray}
E_{kin}~=~\frac{1}{2}M_{CME}V_{bulk}^{2}.
\end{eqnarray}
The valuse are calculated to be of the order of $10^{33}$--$10^{34}$~erg, which are also consistent with other stellar cases. The specific values of the masses (M$_{CME}$) and kinetic energies (E$_{kin}$) can be found in Table~\ref{tab3}.

The S/N of the spectra strongly influences the detectability of stellar CME features. According to \citet{Odert2020}, the minimum detectable CME mass M$_{CME, min}$ could be estimated for the flare from
\begin{eqnarray}
M_{CME, min}~\approx ~\frac{{\pi}R_{\star}^{2}m_{H}N_{H}}{S/N~\times~W[1-e^{-\tau}]},
\end{eqnarray}
in which R$_{\star}$ is the radius of a host star, m$_{H}$ is the mass of hydrogen atom, N$_{H}$ is the column density of a prominence, W is the geometric dilution factor, and $\tau$ is the optical depth of the H$_{\alpha}$ emission line. Typical values of parameters can be found in \citet{Odert2020}. With the S/Ns of the H$_{\alpha}$ line profiles tabulated in Table~\ref{tab3}, the minimum value of 62 corresponds to a minimum detectable CME mass M$_{CME, min}$ of about $9.5~\times~10^{17}$~g. Thus, the derived masses of the CME candidates lie well above this minimum detectable value for our observations.

\begin{figure*}
\includegraphics[width=9.cm,height=7.5cm]{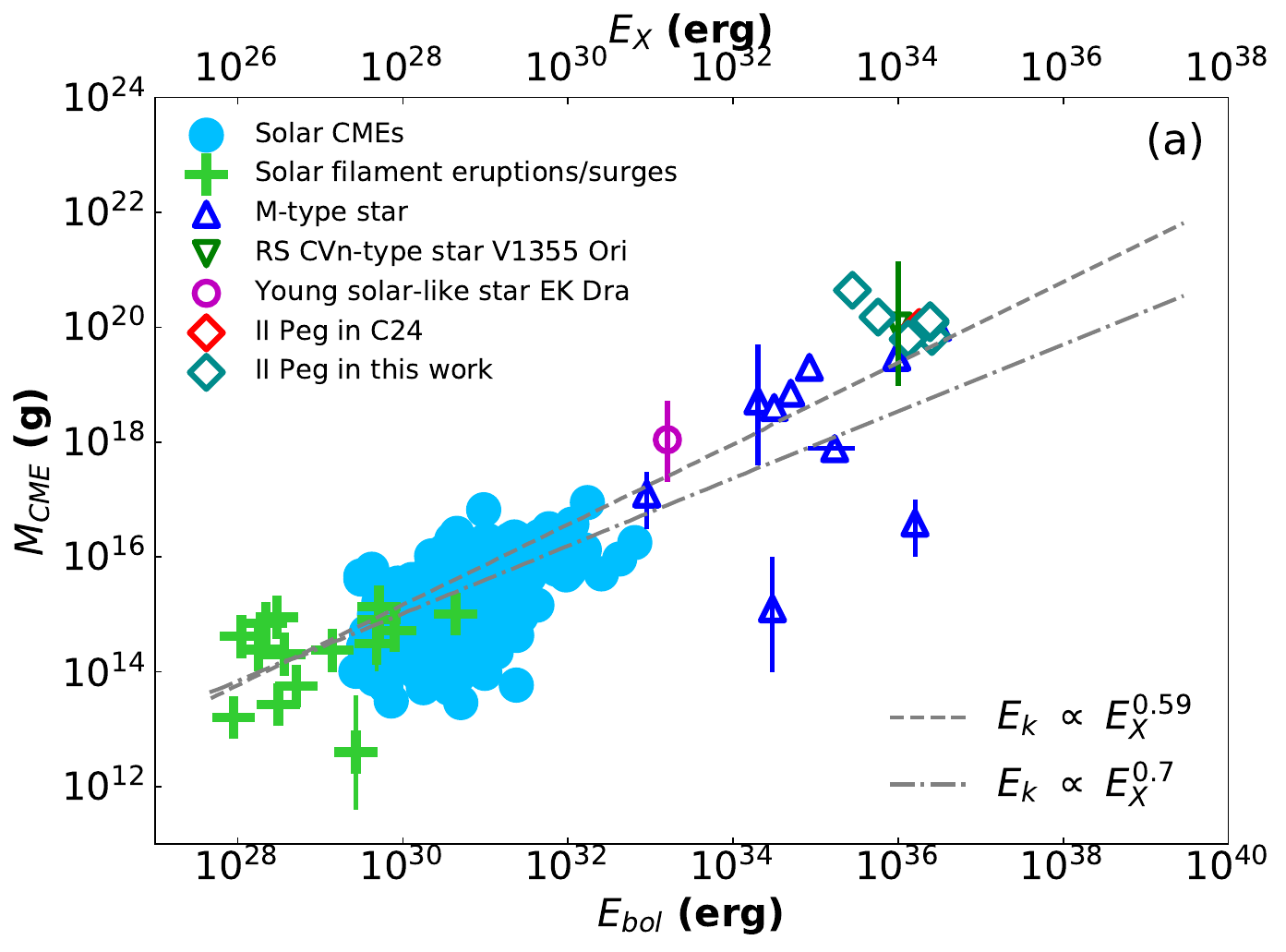}
\includegraphics[width=9.cm,height=7.5cm]{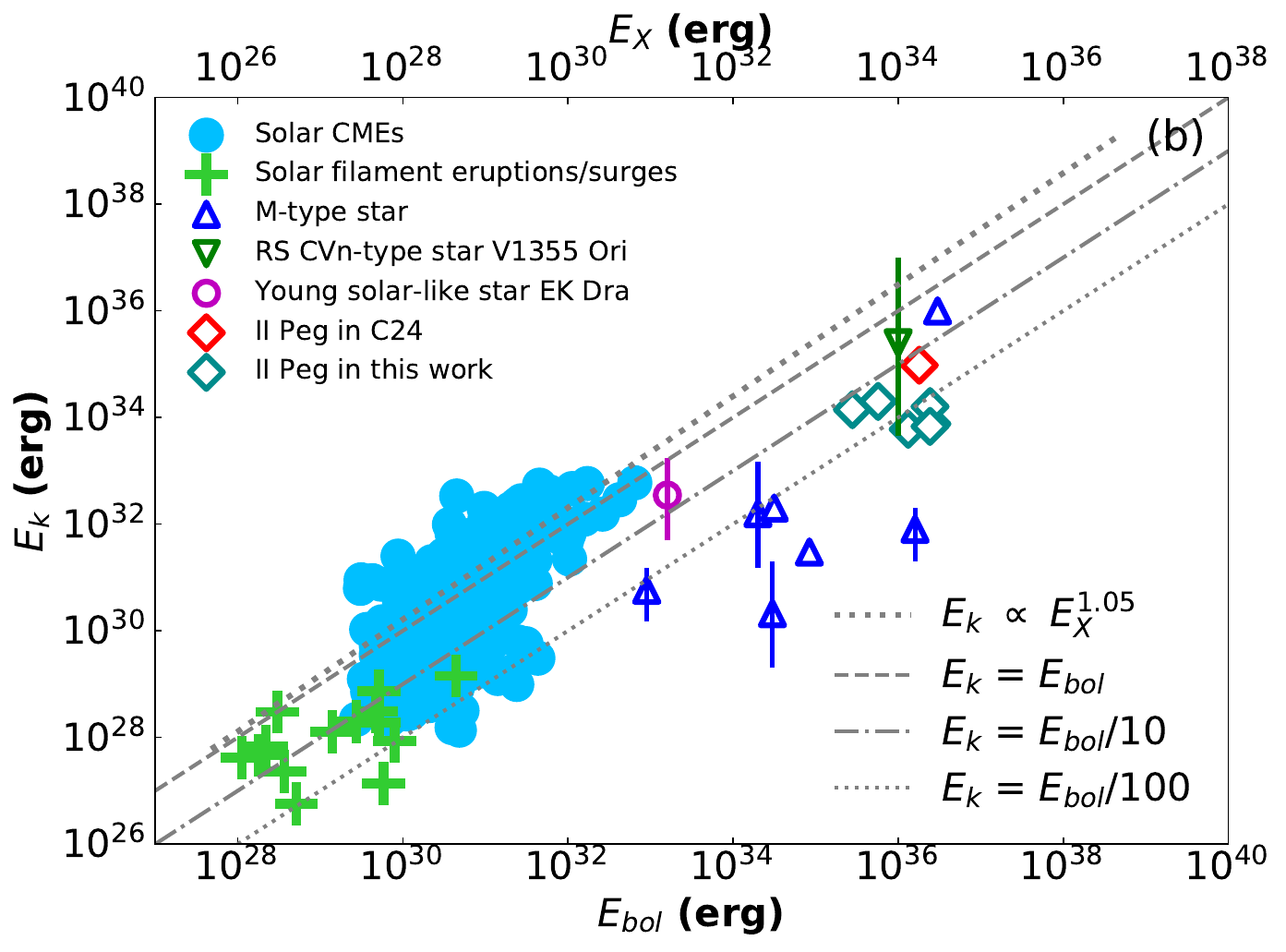}
\caption{Comparison with other events. (a) CME mass $M_{CME}$ and (b) kinetic energy $E_{k}$ are plotted as functions of flaring bolometric energy (bottom horizontal axis) and X-ray energy emitted in the GOES band (top axis). The Solar CME data are derived from \citet{Yashiro2009} and indicated by filled sky-blue circles. The Solar filament eruptions/surges are obtained from \citet{Namekata2024}, which were originally taken from \citet{Namekata2022} and \citet{Kotani2023}, and presented by green plus. The trend fits to the solar CMEs are presented by the red dotted lines expressed as $M_{CME}$~$\propto$~$E_{X}^{0.59}$ and $E_{k}$~$\propto$~$E^{1.05}$ \citep{Drake2013} , and the black dashed line expressed as $M_{CME}$~$\propto$~$E_{X}^{0.7}$ \citep{Aarnio2012}. Similar as \citet{Namekata2024}, the relations of $E_{k}$~=~$E_{bol}$, $E_{k}$~=~$E_{bol}$/10, and $E_{k}$~=~$E_{bol}$/100 are also plotted in the right panel. Stellar CME candidates on different type stars are obtained from \citet{Moschou2019}, \citet{Wang2021, Wang2022}, \citet{Namekata2022}, and \citet{Inoue2023}.}
\label{Fig13}
\end{figure*}

The masses of CME candidates are two to three orders of magnitude greater than the most massive solar CME, which has a mass of approximately $10^{17}$~g \citep{Gopalswamy2009}. In contrast, the kinetic energy values associated with these candidates are only one order of magnitude higher than the maximum kinetic energy observed in solar CMEs, estimated at around $10^{33}$~erg \citep{Gopalswamy2009}, or they may be comparable in magnitude. Moreover, to compare the detected CME candidates with other events, including solar CMEs, solar prominence/filament eruptions/surges, and CME candidates on the other stars, we plot their masses and kinetic energies as functions of the flare energy emitted in the GOES X-ray and bolometric flare energy in Figure~\ref{Fig13}. The flare energies emitted in the H$_{\alpha}$ line are converted to the GOES X-ray (1--8~\AA~band) wavelength by using the scaling relation between H$_{\alpha}$ and GOES soft X-ray flare energies in Figure~2 and Equation~(1) of \citet{Haisch1989}, and then we use a correction of $E_{bol}$ = 100$E_{X}$ \citep{Emslie2012} to obtain the bolometric flare energy. On the one hand, it can be seen from the figure that the mass of CME candidate observed on 2017 December~5  greatly exceeds the predictions of the solar flare-CME relation, while the values of the other candidates are very close to the relation of $M_{CME}$~$\propto$~$E_{X}^{0.7}$ derived by \citet{Aarnio2012}. The distribution of masses of CME candidates on stars is observed to be scattered, yet is generally consistent with the solar flare-CME relation. On the other hand, similar to solar prominence/filament eruptions/surges and CME candidates on the other stars, the kinetic energies of CME candidates on II~Peg are less than the one extrapolated from solar CMEs and the values obtained in this work are consistent with the relation of $E_{k}$~=~$E_{bol}$/100. The primary reason for the lower kinetic energies is that prominence eruptions on stars generally have much smaller velocities than CMEs on the Sun \citep{Maehara2021, Namekata2022}, just as solar prominence/filament eruptions/surges have lower kinetic energies. Moreover, the overlying magnetic field on active stars might reduce the CME speeds and therefore results in the small kinetic energies \citep[e.g.][]{Drake2016, Alvarado2018}.

\subsubsection{A special case in the CME candidates}
Among these CME candidates hunted on II~Peg, there is one particular instance. For the CME candidate observed on 2017 December~5, it is the most massive event observed on II~Peg and the only one detected by using the different chromospheric activity lines (including the H$_{\alpha}$, H$_{\beta}$, and $\mbox{He~{\sc i}}$~D$_{3}$ lines, see Section~4.3) formed at different temperatures. As stated by \citet{Giampapa1978}, the $\mbox{He~{\sc i}}$~D$_{3}$ line emission typically occurs at temperature around 20,000~K. Therefore, the blueshifted broad emission of the $\mbox{He~{\sc i}}$~D$_{3}$ line indicates the presence of plasma with temperature exceeding 20,000~K in the erupted materials. However, no analogous emission feature is observed in the $\mbox{He~{\sc ii}}$~4686~\AA~line. This is due to the fact that the $\mbox{He~{\sc ii}}$~4686~\AA~line emission occurs at very high temperature $\textgreater$~30,000~K \citep{Zirin1988, Lamzin1989}, which suggests that plasma with such high temperature is absent in the erupted prominence materials.

In addition, as previously discussed in Section 6.1.1, the blueshifted emission component could also be attributed to chromospheric evaporation. Nevertheless, the hypothesis of chromospheric evaporation might be dismissed with regard to this blueshifted emission component. The rapid expansion of heated materials from the chromosphere to the corona during chromospheric evaporation has been observed to result in the formation of spectral lines at different temperatures \citep[e.g.][]{Tei2018, Li2023}. Therefore, the observational characteristic of the appearance of the blueshifted emission component in the cooler $\mbox{He~{\sc i}}$~D$_{3}$ line and the absence in the hotter $\mbox{He~{\sc ii}}$~4686~\AA~line indicates that the blueshifted emission component cannot be attributed to a chromospheric evaporation. In other words, the most likely cause of the blue-shifted emission component observed on 2017 December~5, is a prominence eruption, confirmed by the spectral lines at different formation temperatures, which has the potential to develop into a CME, as discussed in Section 6.1.1.

\subsubsection{CME occurrence rate}
\citet{Muheki2020a, Muheki2020b} analyzed H$_{\alpha}$ line asymmetry using long-term high-resolution spectroscopy of the active M-type stars AD~Leo and EV~Lac. Some asymetric signatures were found in the line profile of flaring spectra. In addition to obtaining spectroscopic observations of one star, some research works have been conducted to detect possible stellar CMEs by investigating a large number of stars based on archival data \citep[e.g.][]{Fuhrmeister2018, Vida2019, Leitzinger2020, Koller2021, Lu2022}. However, only a few of CME candidates were detected in these works. A number of potential reasons for the difficulty in detecting stellar CMEs have been discussed, including observational biases, lower activity levels, and magnetic suppression \citep{Alvarado2018, Leitzinger2020, Koller2021}.

Over the course of the long-term high-resolution spectroscopic observations from 1999 to 2022, we had collected 348 spectra of II Peg. The total monitoring time was about 188 hours which we got by adding up the exposure times of all the spectra. We have identified more than 19~optical flare events in total and 7~flares are presented in this study due to asymmetric features in the H$_{\alpha}$ line profiles. This suggests that approximate 0.1~flares occur per hour on II~Peg. The bolometric white-light energies of all flares are in the range of $\ga~10^{35}$~erg, which are larger than the most energetic solar flares. Six flare associated CME candidates are found in this paper and one was presented in C24. Therefore, the possible flare--CME association rate is about 30\% on II~Peg, which is also much smaller than the association rate for the most energetic flares (approximate~100\%) on the Sun \citep[e.g.][]{Yashiro2006,Aarnio2011}. To investigate the cause of the low flare-CME association rate on II~Peg, it may be necessary to first remove observational biases, for example, by improving temporal resolution of flare observations in the future studies.

\subsection{Redshifted extra absorption signatures}
Except the Doppler-shifted emission signatures, there are some redshifted excess absorptions in the subtracted H$_{\alpha}$ line profiles during our observations. Similar redshifted excess absorptions have also been observed in the H$_{\alpha}$ line profiles on RS CVn-type binary stars IM~Peg and $\sigma$~Gem \citep{Cao2022}, which are interpreted as due to downflow of cool absorbing materials on the stellar disk. Moreover, \citet{Honda2018} found an absorption component with a velocity of a few tens km~s$^{-1}$ in the red wing of the H$_{\alpha}$ line in the early and later phases of the flare on M-dwarf EV~Lac and thought the absorption feature might be caused by plasma downflows produced in the post-flare loops. For our situation, as presented in Section~\ref{sec5}, the redshifted excess absorption signatures could be observed not only during the flares but also in the spectra without flare. 

For our situation, we attribute the redshifted absorption signatures for both cases (with and without flares) to coronal rain. Coronal rain is a well-known and common observed phenomenon in the solar atmosphere, in which dense and cool plasma condensation forms in the hot corona, and then falls to the solar surface along magnetic loops \citep{Antolin2012}. Depending on the relation with flare, coronal rain is usually classified into two categories: flare-driven rain and quiescent rain \citep{Antolin2020}. The flare-driven coronal rain frequently appears in post-flare loops, as mentioned in Section~6.1.1, while the quiescent rain corresponds to a more ubiquitously observed rain in active region loops. In Section 6.1.1, the typical velocity of coronal rain is presented in the solar case. The difference is that we discussed the redshifted emission signatures previously, here we refer to the absorption signatures. Similar to prominences on the Sun and in the stellar cases, coronal rains can be observed as emission out of the solar/stellar limb and absorption on the disk, respectivly. Moreover, during our observations of II~Peg, the redshifted excess absorption signatures with bulk velocities ranging from 76 to 143 ~km~s$^{-1}$ (see Table~\ref{tab4}) are consistent with the values of coronal rains on the Sun. Furthermore, it is notable that the velocities of the extra absorption features associated with the flares are larger than the ones in the cases without flares, in consistent with the solar observations.
\section{Summary and conclusions}\label{sec7}
II~Peg is a RS~CVn-type star that is highly active and frequently produces large flares. Due to the close connection between highly energetic flares and CMEs on the Sun, II~Peg may have frequent CME occurrences. Based on long-term high-resolution spectroscopic monitoring dataset, we attempt to detect the possible stellar prominence eruptions and CMEs on II~Peg by analyzing asymmetric features of the chromospheric lines H$_{\alpha}$ and H$_{\beta}$ during the flares.

We have identified seven optical flare events characterized by asymmetric spectral profiles and higher energy releases. The low limits of the flare energies are of the order of $10^{33}$--$10^{35}$~erg in the H$_{\alpha}$ line. The bolometric white light flare energies are derived to be $10^{35}$--$10^{36}$~erg, which suggest that these flares are comparable to stellar superflares. 

Five of these flare events were associated with six potential CMEs, which were detected by the Doppler-shifted eimission signatures in the chromospheric line profiles, especially the H$_{\alpha}$ line. The Doppler-shifted emission signatures include two redshifted emission components and four blueshifted emission ones. The masses of CME candidates are estimated to be in the range of $10^{19}$--$10^{20}$~g, while the kinetic energies are calculated to in the range of $10^{33}$--$10^{34}$~erg. The masses are two to three orders of magnitude larger than the most massive solar CME, while their kinetic energies are either one order of magnitude larger than the maximum kinetic energy of solar CMEs or they are comparable in magnitude. The most massive CME candidate was simultaneously observed in the H$_{\alpha}$, H$_{\beta}$, and $\mbox{He~{\sc i}}$~D$_{3}$ lines. For II~Peg, the possible flare-CME association rate is about 30\%. Additionally, it is important to note that the Doppler-shifted emission signatures may also be caused by other forms of material motions during the flares, such as chromospheric evaporation and condensation, as well as the falling materials in prominence eruptions.

Apart from the Doppler-shifted emission components, there are some red-shifted excess absorption signatures in the subtracted H$_{\alpha}$ line profiles. These excess absorption features were observed not only during the flares but also in spectra without flare, which may be interpreted as originating from flare-driven coronal rain and quiescent rain, respectively.

\begin{acknowledgments}
We are grateful to the referee's helpful suggestions and comments, which led to improvements of the manuscript. This work is supported by the National Natural Science Foundation of China under grant No. 12288102. We acknowledge the support of the staff of the Xinglong 2.16m telescope. This work is partially supported by National Astronomical Observatories, Chinese Academy of Sciences. The present study is also financially supported by the National Natural Science Foundation of China under grant Nos. 10373023, 10773027, and U1531121, the Yunnan Fundamental Research Projects (grant Nos.~202201AT070186 and 202305AS350009), the Yunnan Revitalization Talent Support Program (Young Talent Project), International Centre of Supernovae, Yunnan Key Laboratory (No.~202302AN360001), and the China Manned Space Program with grant No. CMS-CSST-2025-A15.
\end{acknowledgments}

\bibliography{sample631}{}
\bibliographystyle{aasjournal}

\end{document}